\documentclass[12pt]{article}

\usepackage{graphicx}
\usepackage{amsfonts}

\def\gtwid{\mathrel{\raise.3ex\hbox{$>$\kern-.75em\lower1ex\hbox{$\sim$}}}}
\def\ltwid{\mathrel{\raise.3ex\hbox{$<$\kern-.75em\lower1ex\hbox{$\sim$}}}}
\def\square{\kern1pt\vbox{\hrule height 1.2pt\hbox{\vrule width 1.2pt\hskip 3pt
   \vbox{\vskip 6pt}\hskip 3pt\vrule width 0.6pt}\hrule height 0.6pt}\kern1pt}

\begin{document}

\begin{titlepage}

\begin{flushright}
UFIFT-QG-15-01
\end{flushright}

\vskip 1cm

\begin{center}
{\bf Strange Work in Strange Places:\\
Quantum Field Theory in Curved Space}
\end{center}

\vskip 1cm

\begin{center}
S. P. Miao$^{1*}$ and R. P. Woodard$^{2\dagger}$
\end{center}

\vskip .5cm

\begin{center}
\it{$^{1}$ Department of Physics, National Cheng Kung University \\
No. 1, University Road, Tainan City 70101, TAIWAN}
\end{center}

\begin{center}
\it{$^{2}$ Department of Physics, University of Florida,\\
Gainesville, FL 32611, UNITED STATES}
\end{center}

\vspace{.5cm}

\begin{center}
ABSTRACT
\end{center}

Astronomers seem to be observing the fossilized remnants of 
quantum gravitational processes which took place during the 
epoch of {\it primordial inflation} that is conjectured to
have occurred during the first $10^{-32}$ seconds of cosmic
history. We give a non-technical description of what causes
these processes and how they become preserved to survive to
the current epoch. We also discuss some of the secondary 
effects which should result.

\begin{flushleft}
\end{flushleft}

\vskip 1cm

\begin{flushleft}
$^{*}$ e-mail: spmiao5@mail.ncku.edu.tw \\
$^{\dagger}$ e-mail: woodard@phys.ufl.edu
\end{flushleft}

\end{titlepage}

\section{Introduction}\label{intro}

There are two reasons why quantum field theory in curved space might 
not be considered an appropriate topic for a book celebrating the 
100th anniversary of general relativity:
\begin{enumerate}
\item{Albert Einstein never accepted quantum mechanics; and}
\item{Applying quantum mechanics to gravity may show that general
relativity is not the correct theory of gravity.}
\end{enumerate}
However, Einstein loved to pursue ideas to their logical conclusions,
and he did important work on quantum mechanics despite his misgivings
about it. So the research we will describe really is in the tradition of 
the great man's thought, and it certainly concerns his theory of general 
relativity.

\begin{figure}[htp]
\centering{\includegraphics[scale = 0.6]{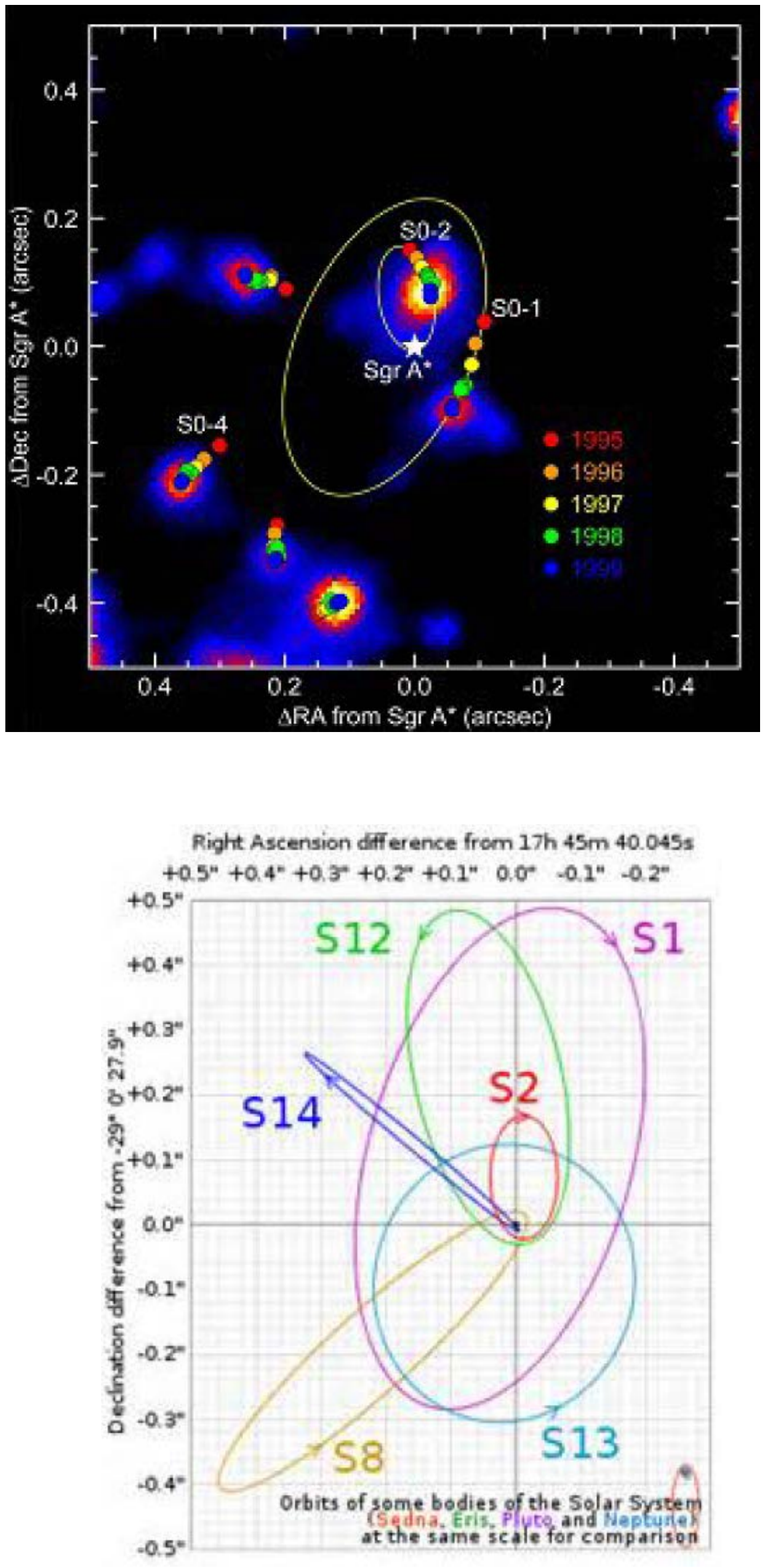}}
\caption{Stars orbiting the giant black hole near the center of our 
galaxy in the constellation of Sagittarius.}
\label{black}
\end{figure}

Until recently, most work on quantum field theory in curved space was 
motivated by events from the 1970's. This fertile decade witnessed 
Stephen Hawking's brilliant insight that quantum mechanics causes black
holes to emit thermal radiation \cite{Hawking:1974sw}, and William 
Unruh's demonstration that accelerated observers should experience a 
similar effect even in flat space \cite{Unruh:1976db}. These are results 
of the first magnitude, and still an area of active research as regards 
the eventual decay of black holes. However, all of the calculations of
this sort which it is presently possible to perform seem to have been 
done. There are plenty of things we still don't understand, but 
answering those questions is likely to require measurements of the 
effects to motivate simplifying assumptions that would facilitate more
ambitious computations. Unfortunately, no such measurements exist, nor 
are any likely to become available in the near future because the known 
black hole candidates are all large (which means the radiation they 
emit is small) and very far away. For example, Figure~\ref{black} shows 
the orbits of star around the giant black hole near the center of our 
galaxy. From these orbits we can estimate its mass to be about $4.3$ 
million times the mass of our own Sun. The radiation coming from a black 
hole of that size would have a temperature only about $10^{-14}$ degree 
Kelvin above absolute zero. We could not detect this faint a source of 
radiation even it were produced in an Earth-bound laboratory, and it is 
in fact about $2.5 \times 10^{17}~{\rm km}$ away. There are smaller 
black hole candidates which have higher temperatures and are nearer, but 
the numbers are still impossible.

\begin{figure}[htp]
\centering{\includegraphics[scale = 0.09]{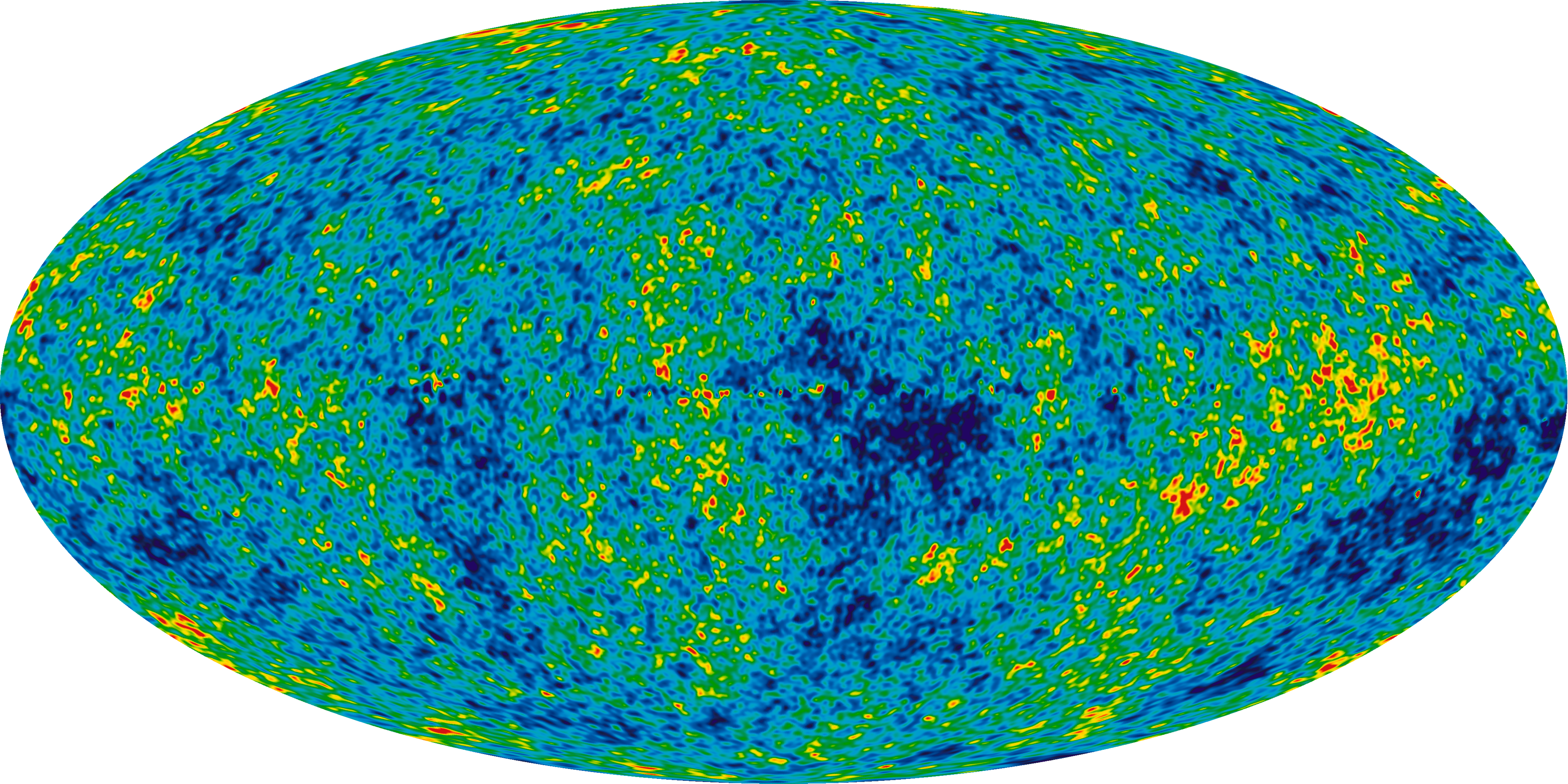}}
\caption{Relative temperatures (cooler for blue, warmer for red) of
radiation from different regions of the sky as imaged by the Planck
satellite. The average temperature is $T_0 = 2.7255$ degrees on the
Kelvin scale, and the fluctuations are about $3 \times 10^{-5}$ 
degree Kelvin.}
\label{Planck}
\end{figure}

Recent years have seen much work on quantum field theory effects in
cosmology. Unlike the situation for black holes, there are still plenty
of relatively simple calculations to be made. There is also an abundant
stream of data to test the results, and to motivate improvements. For
example, Figure~\ref{Planck} displays a full sky map from the Planck 
satellite of the tiny temperature fluctuations which were impressed upon 
radiation from the Big Bang by the gravity from quantum fluctuations in 
matter during the first $10^{-32}$ seconds of existence. These same 
fluctuations served as the seeds for the aggregation of matter to form
stars and galaxies. Among other things, they represent the first quantum 
gravitational effects which have ever been resolved. About $10^8$ 
bits of this data have already been recovered. The present technique for
resolving these fluctuations gives us something like an x-ray of the
universe because it reveals the direction in which the fluctuation is 
located but not its distance from us. There is a more promising method 
in which one measures the redshift of the 21 centimeter radiation emitted 
by Hydrogen gas throughout the universe. Greater redshift means further 
distance so we can reconstruct what is effectively a CT scan of the 
universe, containing vastly more information. The full development of 
this technique over the course of the next several decades might provide 
a staggering $10^{21}$ bits of data \cite{Loeb:2003ya}. These are the
sorts of numbers which intoxicate physicists.

So topicality, simplicity and ready access to data all combine to
draw attention away from quantum effects in black holes to quantum 
effects in cosmology. There is also the important consideration of
the present authors' expertise. As interesting as we find the work 
on black holes, it is not our tale to tell, whereas we have worked 
extensively on quantum field theory in cosmology.

Ours is not a simple story. Men of genius have laboured for centuries 
to extend human understanding of natural law. The magnitude of their 
achievement is evident from the huge barrier of background information
which must be assimilated before accounts of modern physics become 
comprehensible. Even graduate students in our subject are dismayed by 
the frequency with which the dismissive phrase "Introduction to" 
appears in the titles of their courses and textbooks! We have done our
best to distil the essential facts and explain them in section 2. 
Section 3 describes the production of scalars and gravitons in 
cosmology, which is thought to have produced the fluctuations imaged
in Figure~\ref{Planck}. These might be termed the primary effect of
cosmological expansion. Section 4 describes some of the fascinating
secondary effects. In section 5 we list some of the problems which are
still open.

We close with a note of caution. The general public is rightly impressed 
with the erudition displayed by physicists and astronomers, but one should 
bear in mind that they are still human, and as liable as anyone else to
harbour unreasonable prejudices. In the panel discussion at a 1997 
conference on the mounting evidence for the existence of black holes, 
Werner Israel shared a charming anecdote about this \cite{WI1}:
\begin{quotation}
Twenty years ago I spent a very pleasant sabbatical year at the $\ldots$
Institute in $\ldots$ . Although this story is true, I have blanked out 
the names because its significance is generic -- it could have happened
at any of dozens of institutions at that time. Shortly after my arrival 
there was a coffee party, and after some warm words of welcome, the Director
of the Institute remarked, ``Werner is going to be with us for a year. We
should all talk to him and try to cure him of these silly notions he has 
about the possibility of black holes.''
\end{quotation}
The Schwarzschild solution was discovered in 1916, just one year after 
the introduction of general relativity. And it was obvious, from the 
recognition of white dwarf stars in the 1920's, that compact objects very
close to black holes exist in nature \cite{WI2}. (Indeed, our Sun will end
this way.) Yet learned physicists for decades rejected the possibility that 
slightly more massive stars might collapse to form black holes. While this
attitude had softened to bemused toleration by Israel's time, the future
Nobel laureate Subrahmanyan Chandrasekhar was publicly ridiculed during the
1930's for his demonstration that white dwarf stars of more than 1.4 solar 
masses cannot avoid collapse. The history of science is littered with 
similar cases, and it leads to dismal reflections on the human condition 
that no generation of scientists ever seems to rise above the reflex to 
persecute ideas which they find unattractive. So we enjoin the public to 
scepticism about unsupported opinion, no matter who espouses it. Scientists 
should be trusted only as far as their pronouncements are confirmed by 
experiment and observation, or by a provable chain of deduction from 
principles which have been established by experiment and observation. 

\section{Some Key Facts}\label{key}

This section is intended to give simplified explanations of a number of 
key physics theories and techniques, and we begin with a review of notation.
Large and small numbers are represented using ``scientific notation,'' for
example,
\begin{equation}
7.28 \times 10^{-27} \qquad {\rm and} \qquad 3.085 \times 10^{16} \; .
\end{equation}
We use the MKS system of units in which mass is measured in kilograms (kg),
length in meters (m), time in seconds (s) and charge in Coulombs (C). Five 
physical constants have standard symbols:
\begin{eqnarray}
{\rm Speed\ of\ light} & \equiv & c \approx 3.00 \times 10^{8} 
\, \frac{\rm m}{\rm s} \; , \\
{\rm Newton's\ Constant} & \equiv & G \approx 6.67 \times 10^{-11} 
\, \frac{{\rm m}^3}{\rm kg\!\!-\!\!s} \; , \\
{\rm Reduced\ Planck\ Constant} & \equiv & \hbar \approx 1.054 \times 
10^{-34} \, \frac{{\rm kg\!\!-\!\!m}^2}{\rm s} \; , \\
{\rm Proton\ Charge} & \equiv & e \approx 1.602 \times 10^{-19} \, 
{\rm C} \; , \\
{\rm Electric\ Permittivity\ of\ free\ space} & \equiv & \epsilon_0 
\approx \times 10^{-12} \, \frac{{\rm C}^2\!\!-\!\!{\rm s}^2}{{\rm 
kg\!\!-\!\!m}^3} \; . 
\end{eqnarray}
Note that we use triple horizontal lines $(\equiv)$ to indicate the
quantity on the left ``is defined to be'' the quantity on the right. 
Scientists reserve the equals sign ($=$) for cases in which both sides
of the relation have independent definitions.
 
In both physics and mathematics we have tried to employ concepts from 
high school which should be familiar to any educated reader. However, 
calculus was invented to give the laws of physics their simplest 
expression and there are some points at which it cannot be avoided. 
We use an over-dot to indicate differentiation with respect to time,
\begin{equation}
\dot{f}(t) \equiv \frac{df(t)}{dt} \qquad , \qquad \ddot{f}(t)
\equiv \frac{d^2f(t)}{dt^2} \; .
\end{equation}
Some more technical discussions have been relegated to appendices.
Although we will not employ any vector or tensor calculus, we will 
use vectors. They are indicated by an over-arrow,
\begin{equation}
\vec{r} \equiv (x,y,z) \qquad , \qquad \vec{A} \equiv (A_x,A_y,A_z) \; .
\end{equation}
The scalar product of two vectors is indicated with a dot,
\begin{equation}
\vec{A} \!\cdot\! \vec{B} \equiv A_x B_x + A_y B_y + A_z B_z \; .
\end{equation}
We shall also occasionally need complex numbers and their norms,
\begin{equation}
z = a + i b \qquad \Longrightarrow \qquad \vert z\vert^2 = a^2 + b^2 \; .
\end{equation}
An important special case is the exponential of an imaginary number,
\begin{equation}
e^{i \theta} = \cos(\theta) + i \sin(\theta) \; . \label{Euler}
\end{equation}
Remember that the arguments of trigonometric functions such as those
in expression (\ref{Euler}) depend upon radians, not degrees!

\subsection{General Relativity}

Everyone who has studied physics is familiar with Isaac Newton's three
laws of motion. In the original Latin his first law reads: 
\begin{quotation}
Corpus omne perseverare in statu suo quiescendi vel movendi uniformiter 
in directum, nisi quatenus a viribus impressis cogitur statum illum mutare.
\end{quotation}
Which is to say, "Every body persists in its state of being at rest
or of moving uniformly straight forward, except insofar as it is compelled
to change its state by force impressed." Einstein's theory of general
relativity describes how gravity modifies the geometrical concept of a
straight line so that an object moving in a gravitational field, with no
other forces, really does follow a straight line. General relativity does 
this by introducing a new dynamical variable known as the metric tensor 
field, which pervades space and time just like the electromagnetic fields 
which carry electric and magnetic forces and also describe the propagation 
of light.

The metric tensor field quantifies how observers at different points in 
space and time compare times, distances and the directions needed to 
determine if something is moving in a straight line. Popular science 
readers are probably aware of statements about how moving clocks seem 
run more slowly, and how objects seem contracted in the direction of 
motion, with both effects becoming large as the velocity approaches the 
speed of light. Those are special relativistic concepts which describe 
how observers at the same point, but moving with respect to one another, 
compare their notions of time, distance and direction. General relativity 
has to do with how observers at different points make the same comparison, 
even if they are not moving with respect to one another.

Defining a dynamical theory such a general relativity consists of 
specifying three things:
\begin{enumerate}
\item{What the dynamical variable is;}
\item{How the dynamical variable affects the rest of physics; and}
\item{How the rest of physics affects the dynamical variable.}
\end{enumerate}
Because we must avoid complicated mathematics it is best to proceed by 
analogy with electromagnetism which should be familiar from introductory
physics. The dynamical variables of electromagnetism are the electric 
and magnetic fields, and they affect physics through the forces they 
exert on charged particles. These fields are sourced by charges and
currents through Maxwell's equations. We have just seen that the dynamical 
variable of general relativity is the metric tensor field, and that
it affects the rest of physics by defining the way times, lengths and
directions are compared at different points in space and time. It turns
out that the source for the metric tensor field is the energy and momentum 
per unit volume, and the stresses (non-gravitational force per unit area) 
at each point in space and time. In the elegant mathematics of general
relativity these sources are assembled into a single quantity known as the 
stress-energy tensor. In general relativity the stress-energy tensor 
sources the metric tensor field through the Einstein equations.

\begin{figure}[htp]
\centering{\includegraphics[scale = .8]{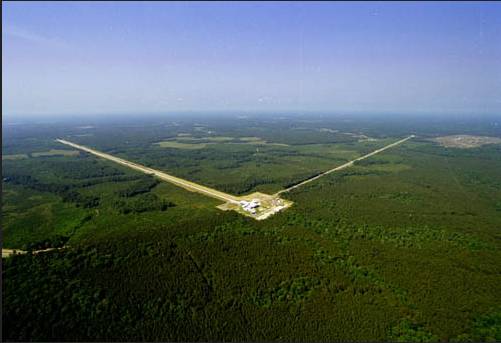}}
\caption{The Laser Interferometer Gravitational Wave Observatory (LIGO) 
located at Livingston, Louisiana in the United States. Another American
LIGO detector exists at Hanford, Washington. These detectors work by using 
laser light bounced back and forth inside a 4 kilometer vacuum pipe.
When a gravitational wave passes through the pipe it induces small 
distortions which can be detected by precise measurements of interference 
patterns produced by the light.}
\label{LIGO}
\end{figure}

No sketch of this brevity can do justice to the scope of general
relativity but our narrative does require a comment on the triune 
nature of general relativity and of all other known force laws. For 
reasons which are not fully understood, the dynamical variables of all
force fields break up into three parts:
\begin{enumerate}
\item{A {\it gauge part} which can be fixed arbitrarily and amounts, for
gravity, to the choice of coordinate system;}
\item{A {\it constrained part} that is determined by the source, which
is the stress-energy tensor for general relativity; and}
\item{A {\it radiation part} which can exist without any source and is
generically produced by making sources undergo acceleration.}
\end{enumerate}
Humans see using electromagnetic radiation; we produce it by causing
charges to accelerate. We have not yet directly detected gravitational 
radiation but there is strong indirect evidence for its existence because 
we can observe the effect of energy being carried away by gravitational 
radiation from pairs of neutron stars which are closely orbiting around 
one another. (Recall that even uniform circular motion entails centripetal
acceleration.) Figure~\ref{LIGO} shows one of the devices which are 
attempting to directly detect gravitational waves. In mid 2015 this
device will resume operation at a greater sensitivity than ever before,
and hopes are high that it will observe the bursts of gravitational 
radiation from mergers between compact objects such as neutron stars
and black holes.

\subsection{Cosmology}

As we stated above, the metric tensor field defines how observers at 
different points in space and time compare times, distances and directions. 
If an observer's time and position vector are $(t,\vec{x})$, then the 
metric tensor field tells him what is the infinitesimal distance to a 
nearby observer whose time and position vector are $(t+dt,\vec{x} +
d\vec{x})$. With no gravitational fields present this relation is,
\begin{equation}
ds^2 = -c^2 dt^2 + d\vec{x} \!\cdot\! d\vec{x} \; . \label{flat}
\end{equation}
Expression (\ref{flat}) encompasses all of special relativity, and 
represents the simplest geometry of general relativity. This way in 
which an earlier theory --- in this case, special relativity --- is 
nested within a more advanced theory --- in this case, general relativity
--- is known as a {\it Correspondence Limit}. All new theories must 
reduce to older theories in the appropriate correspondence limit.

The next most complicated geometry is that of cosmology,
\begin{equation}
ds^2 = -c^2 dt^2 + a^2(t) d\vec{x} \!\cdot\! d\vec{x} \; . \label{FLRW}
\end{equation}  
It differs from the geometry of flat space (\ref{flat}) only by the
function of time $a^2(t)$ which multiplies the space terms. When a
physicist says that the universe is expanding, what he means is that
the scale factor $a(t)$ is growing. It is called ``the scale factor''
because equation (\ref{FLRW}) says that one must multiply coordinate
separations by the scale factor to convert them to physical distance.
We know this expansion is occurring because we can see what it does to
the light emitted (or absorbed) in known atomic transitions whose wave 
length is fixed by the energy of the transition according to physical
laws which we assume to be the same throughout the universe. If this 
light was emitted at time $t$ with wavelength $\lambda$ then the 
wavelength we measure at the current time $t_0$ is,
\begin{equation}
\lambda_0 = \frac{a(t_0)}{a(t)} \times \lambda \equiv (1 + z) \lambda
\; . \label{redshift}
\end{equation}
Equation (\ref{redshift}) assumes that the emitter is not moving 
with respect to us; if it is moving there is also a Doppler shift 
(like the police use to catch speeders!) due to the motion. The factor
$z$ in equation (\ref{redshift}) is known as the (cosmological)
redshift. We can recognize atomic spectral lines which have been 
redshifted by more than a factor of 7, and we can recognize thermal 
radiation which has experienced a redshift of over 1000.

\begin{figure}[htp]
\centering{\includegraphics[scale = 0.5]{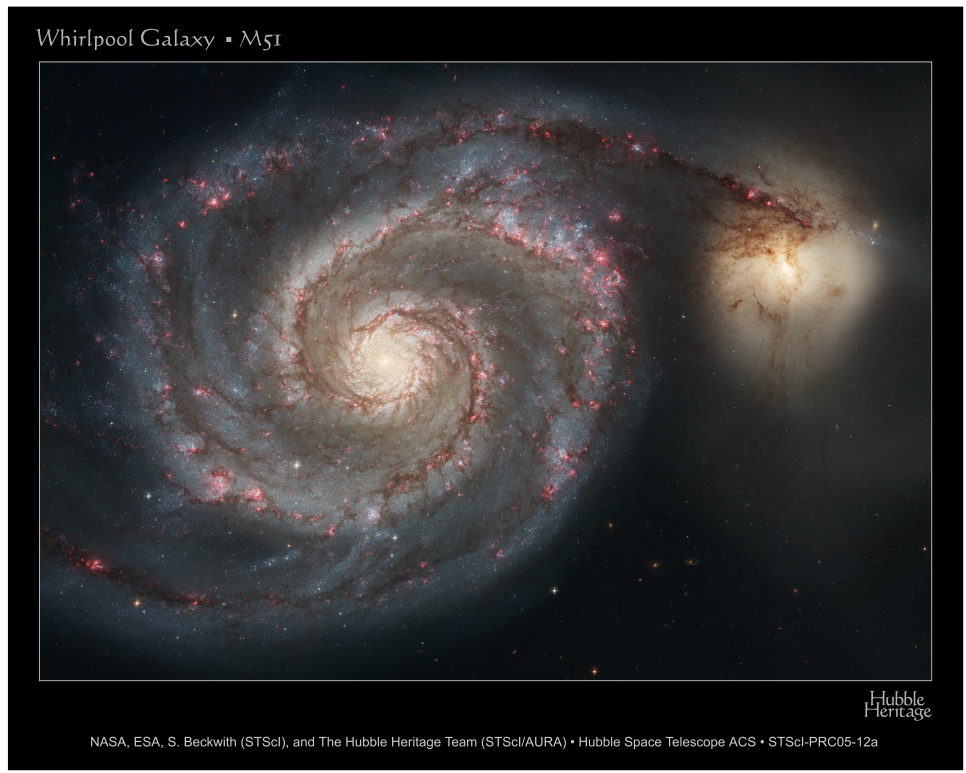}}
\caption{Hubble Space Telescope image of the Whirlpool Galaxy 
(M51), a giant spiral galaxy located about $2.2 \times 10^{23}~{\rm m}$
from us.}
\label{HST}
\end{figure}

Physicists quantify the expansion of the universe through time 
derivatives of the scale factor. The first derivative of its logarithm
is known as the Hubble parameter,
\begin{equation}
H(t) \equiv \frac{d \ln[a(t)]}{dt} = 
\frac{\dot{a}(t)}{a(t)} \; . \label{Hubble}
\end{equation}
Many people have seen the lovely pictures produced by the Hubble Space 
Telescope, for example, Figure~\ref{HST}. However, the actual reason
American tax payers provided a billion US dollars to build and operate 
the telescope was to enable astronomers to measure the current value
of the Hubble parameter to an accuracy of 10\%. Today measurements of
the cosmic microwave background radiation such as Figure~\ref{Planck}
can do an even better job \cite{Planck:2015xua},
\begin{equation}
H_0 = \Bigl(67.74 \pm 0.23\Bigr) \frac{\rm km}{\rm Mpc\!\!-\!\!s} =
\Bigl(2.195 \pm 0.007\Bigr) \times 10^{-18}~{\rm Hz} \; . \label{H0}
\end{equation}
The Hubble parameter was not always so small. The equations of general
relativity predict a relation, known as the $\Lambda$CDM model, which
seems to be valid to redshifts as large as $z \sim 10^{22}$,
\begin{equation}
H = H_0 \sqrt{ \Omega_r (1 \!+\! z)^4 + \Omega_m (1\!+\! z)^3 + 
1 \!-\! \Omega_m \!-\! \Omega_r} \; , \label{Lmodel}
\end{equation}
where observations give the following values \cite{Planck:2015xua},
\begin{equation}
\Omega_r = \Bigl(9.29 \pm 0.050\Bigr) \times 10^{-5} \qquad , \qquad
\Omega_m = 0.315 \pm 0.017 \; .
\end{equation}
The relative abundances of the lightest nuclear isotopes preserve a
fossilized record of conditions at $z \sim 10^9$, during which the
Hubble parameter was 17 orders of magnitude larger than its current 
value. So the picture to keep in mind is a universe which was once
much smaller, much hotter and much more rapidly expanding than it is
now.

\begin{figure}[htp]
\centering{\includegraphics[scale=0.7]{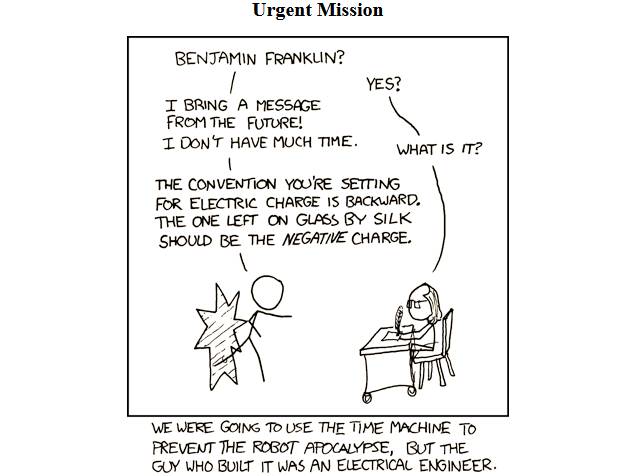}}
\caption{Cartoon about the frustration electrical engineers
feel with Franklin's sign choice.}
\label{Franklin}
\end{figure}

The second derivative of the scale factor is combined with lower 
derivatives to form a dimensionless quantity known as {\it the 
deceleration parameter},
\begin{equation}
q(t) \equiv -\frac{a \ddot{a}}{\dot{a}^2} \; . 
\label{deceleration}
\end{equation}
There is a funny story about the minus sign in expression 
(\ref{deceleration}), which also illustrates the fallibility of 
scientists, the crucial importance of checking beliefs with 
experiment and observation, and just how fast things are changing 
in this particular field. The story begins with modern physicists' 
frustration at Benjamin Franklin's unfortunate choice for the signs 
of electrical charge more than 250 years ago. In those years people 
were just learning about electricity, and they had discovered that 
electricity can be produced by rubbing glass with silk and also by 
rubbing amber with fur; du Fay termed them ``vitreous'' and 
``resinous'' electricity, respectively. Franklin had the brilliant 
insight that these were not two different sorts electricity but 
rather that one represented an excess of charge and the other was a 
relative absence of the same kind of charge. He was quite right 
about that, but he unfortunately chose to label vitreous electricity 
as ``positive'' charge and resinous electricity as ``negative'' charge,
whereas we now understand that rubbing glass with silk carries away
negatively charged electrons and rubbing amber with fur adds 
electrons. Franklin's sign choice means that electrons --- which are 
the basis of our electrical power industry --- carry negative charge. 
One consequence is that electrical engineers must remember that a 
wire carrying positive charge to the right actually consists of 
electrons moving to the left. The cartoon in Figure~\ref{Franklin} 
indicates how irritating they find this!

Physicists hate negative-valued parameters because they inhibit
us from telling the general tendency of things just by glancing at
an equation. Cosmologists were determined not to introduce another 
negative-valued parameter into physics. Because we were certain 
that gravity must be slowing down the expansion --- the same way
it slows a rock which is thrown into the air --- we inserted the minus 
sign into expression (\ref{deceleration}) to keep the parameter positive.
But the universe does as it pleases, without regard to human prejudice,
and its expansion has actually been accelerating for almost 6 billion 
years. So all our cleverness has resulted in a new negative-valued 
parameter \cite{Planck:2015xua},
\begin{equation}
q_0 = -0.54 \pm 0.01 \; .
\end{equation}  
Sometimes you just can't win! This was only shown in 1998, and the 
leaders of the two teams who made the measurement were awarded the 
2011 Nobel Prize in physics. Explaining why the universe is accelerating
is regarded by many as the greatest problem confronting fundamental
theory, and it didn't exist 20 years ago.

Popular science readers have probably seen accounts of how superstring 
theory will provide the ultimate explanation of all natural phenomena. 
It is sobering to quote the words of a leading string theorist when he 
was told the result about cosmic acceleration:
\begin{quotation}
I'm sure the data is wrong because string theory predicts a negative
cosmological constant. And if it's right, I'm going to stop doing 
physics. 
\end{quotation}
Superstring theory {\it does} predict that the final state of the 
universe cannot be one of accelerated expansion, but the observations
showing that our 13.8 billion year old universe is accelerating {\it 
did} turn out to be right. The reaction of string theorists was to
introduce a huge number of free parameters (estimates involve numbers
like $10^{500}$) so that it can be made to produce a long phase of 
accelerated expansion before the ultimate decay into deceleration. It 
can also be made to predict anything else. Scientists have for
generations argued that the hallmark of a scientific theory is its
ability to make predictions which can be checked. The present version
of superstring theory does not meet this criterion but some thinkers
have argued that, rather than abandon string theory, we must change
the definition of ``science.'' We leave the reader to form his own 
opinion about this but we again urge scepticism about pronouncements 
which are not supported by observation and experiment, or by a 
rigorous chain of deduction from principles which have been established 
by observation and experiment.

\begin{figure}[htp]
\centering{\includegraphics[scale=0.7]{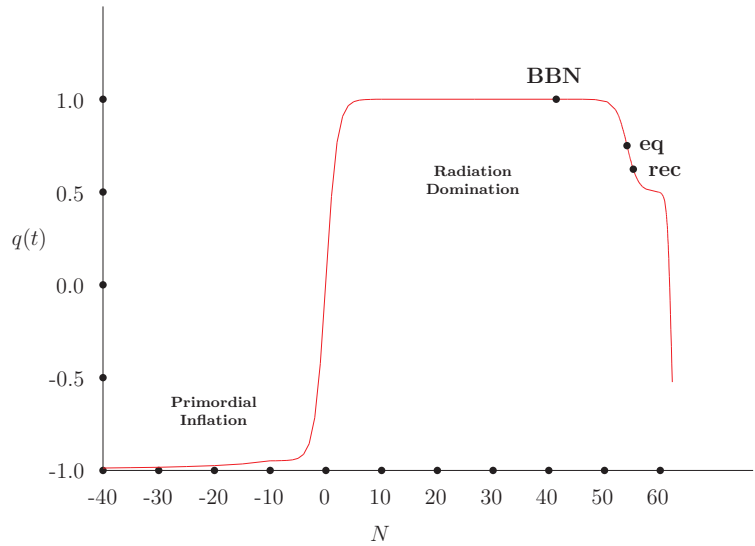}}
\vskip -10cm
\caption{Sketch of how the deceleration parameter is thought to have
changed throughout cosmic history. The parameter $N$ is the number
of e-foldings since the end of primordial inflation. If the scale 
factor at the end of inflation is $a_i$ then $N \equiv \ln[a(t)/a_i]$.}
\label{decel}
\end{figure}

The same $\Lambda$CDM model which led to relation (\ref{Lmodel}) for 
the Hubble parameter also predicts that the deceleration at redshift 
$z$ is,
\begin{equation}
q = \frac{\Omega_r (1 \!+\!z)^4 + \frac12 \Omega_m (1\!+\!z)^3 -1 \!+\!
\Omega_m \!+\! \Omega_r}{\Omega_r (1 \!+\! z)^4 + \Omega_r (1 \!+\! z)^3
+ 1 \!-\! \Omega_m \!-\! \Omega_r} \; . \label{qvsz}
\end{equation}
Even though the universe has been accelerating for almost half of its
life, we can see back to early times during which it was decelerating.
Relation (\ref{qvsz}) implies that the cosmological redshift was only 
about $z \sim .63$ then, which is well within our ability to observe.
This phase of deceleration is important in explaining today's universe
because acceleration works against the usual tendency of gravity to
pull things together. The earlier phase of deceleration is why matter
was able to collapse into stars, then into galaxies and finally into
the galactic clusters we can observe today. After the onset of 
acceleration the formation of larger structures stopped and now the
universe is being pulled apart. Of the approximately 100 billion 
galaxies we can now observe it has been calculated that we will
eventually only be able to see the handful which are gravitationally
bound to our own galaxy.

Considerations of this sort lead one to wonder how the very large scale
universe got to be as uniform as it seems to be. In view of the smoothing
out process which is even now being caused by the current phase of
cosmic expansion, one might suspect that another, much earlier phase
of accelerated expansion could do the job. This epoch is known as
{\it primordial inflation} and it is conjectured to have occurred
during the first $10^{-32}$ seconds of cosmic history. There are very 
strong reasons for taking this idea seriously. One example is the cosmic
microwave background radiation whose tiny temperature fluctuations are 
imaged in Figure~\ref{Planck}. This radiation is left over from a time
about 380,000 years into cosmic history, when the universe was slightly 
over 1000 times smaller than it is today, and hence over 1000 times 
hotter. At that instant the universe finally became cool enough for 
electrons and protons to form neutral Hydrogen, at which point light 
could propagate almost freely. We are seeing the highly redshifted glow 
from that hot plasma. It is in almost perfect thermal equilibrium --- 
the fluctuations shown Figure~\ref{Planck} are only about one part in 
$10^5$, which is far, far better equilibrium than the air of any 
normal room. As everyone knows who has watched ice melt in a glass of 
water, it takes time for things to reach the same temperature after 
they are put in contact. Without primordial inflation the light we 
are seeing would have come from about 3000 regions which are so distant 
from one another that not even light could have travelled between them 
during the brief time the universe had existed by then. Primordial 
inflation avoids the problem by causing all of the sky we can now 
observe to have come from a region which was so small that it had 
plenty of time to reach equilibrium.

Figure~\ref{decel} shows how we think the deceleration parameter
behaved over cosmic history. In order to avoid compressing all early
events into a single point, the scale of time has been replaced by
$N$, the number of e-foldings since the end of inflation. If 
primordial inflation ended at time $t_i$ then the value of $N$ at 
any time $t$ is,
\begin{equation}
N = \ln\Bigl[ \frac{a(t)}{a(t_i)}\Bigr] \; .
\end{equation}
Significant events on the graph are:
\begin{itemize}
\item{{\it Primordial Inflation} during which the deceleration 
parameter was very close to $q= -1$ and the Hubble parameter may 
have been 56 orders of magnitude larger than it is today. Note 
that this period is {\it not} described by the $\Lambda$CDM model
(\ref{Lmodel}) and (\ref{qvsz}).}
\item{{\it Radiation Domination} during which the $\Omega_r 
(1+z)^4$ term was much larger than any other term in equations 
(\ref{Lmodel}) and (\ref{qvsz}). Hence the deceleration parameter 
was close to $q = +1$ and the universe was so hot that thermal 
motion made all matter move at nearly the speed of light.}
\item{{\it Big Bang Nucleosynthesis} during which neutrons and
protons fell out of thermal equilibrium and the seven lightest 
nucler isotopes were formed (the heavier ones formed much later 
during supernovae).}
\item{{\it Matter-Radiation Equality} when the $\Omega_r (1 + z)^4$
and $\Omega_m (1 + z)^3$ terms in equation (\ref{Lmodel}) became 
equal.}
\item{{\it Recombination} during which electrons and protons
formed neutral Hydrogen and light was able to propagate freely.}
\end{itemize}

\subsection{Quantum Mechanics}

Quantum mechanics is not a dynamical theory in the sense of having its 
own dynamical variable, force law and source equation. It is rather a 
procedure, known as ``quantization'', which can be applied to any physical 
theory. So one should not imagine that there are special, quantum field 
equations which are different from those of classical field 
theory.\footnote{Note that we use the term ``classical'' in 
contradistinction to ``quantum'', without regard to whether or not special 
or general relativistic effects are included.} It means precisely the 
same thing to solve these equations quantum mechanically as it does 
classically: one expresses the dynamical variable at any time in terms 
of its values --- and those of all other dynamical variables --- at some 
initial time. 

As a very simple example, consider the dynamical system comprised of a
single point particle of mass $m$ moving freely in one dimension whose 
position at time $t$ is $x(t)$. For that system the initial value 
solution is,
\begin{equation}
x(t) = x_0 + \frac{p_0}{m} t \; , \label{qsol}
\end{equation}
where $x_0$ is the initial position and $p_0$ is the initial momentum.
These two initial values are called {\it conjugate pairs} and every
dynamical variable has them. The solution (\ref{qsol}) is valid in both 
classical physics and in quantum physics; the only difference is what 
$x_0$ and $p_0$ mean. In classical physics they are just numbers --- for 
example, $x_0$ might be five meters and $p_0$ might be minus three 
kilogram-meters per second. Quantization consists of making the initial 
values of conjugate pairs into random numbers whose probability 
distributions are tied to one another. 

\begin{figure}[htp]
\centering{\includegraphics[scale = 0.3]{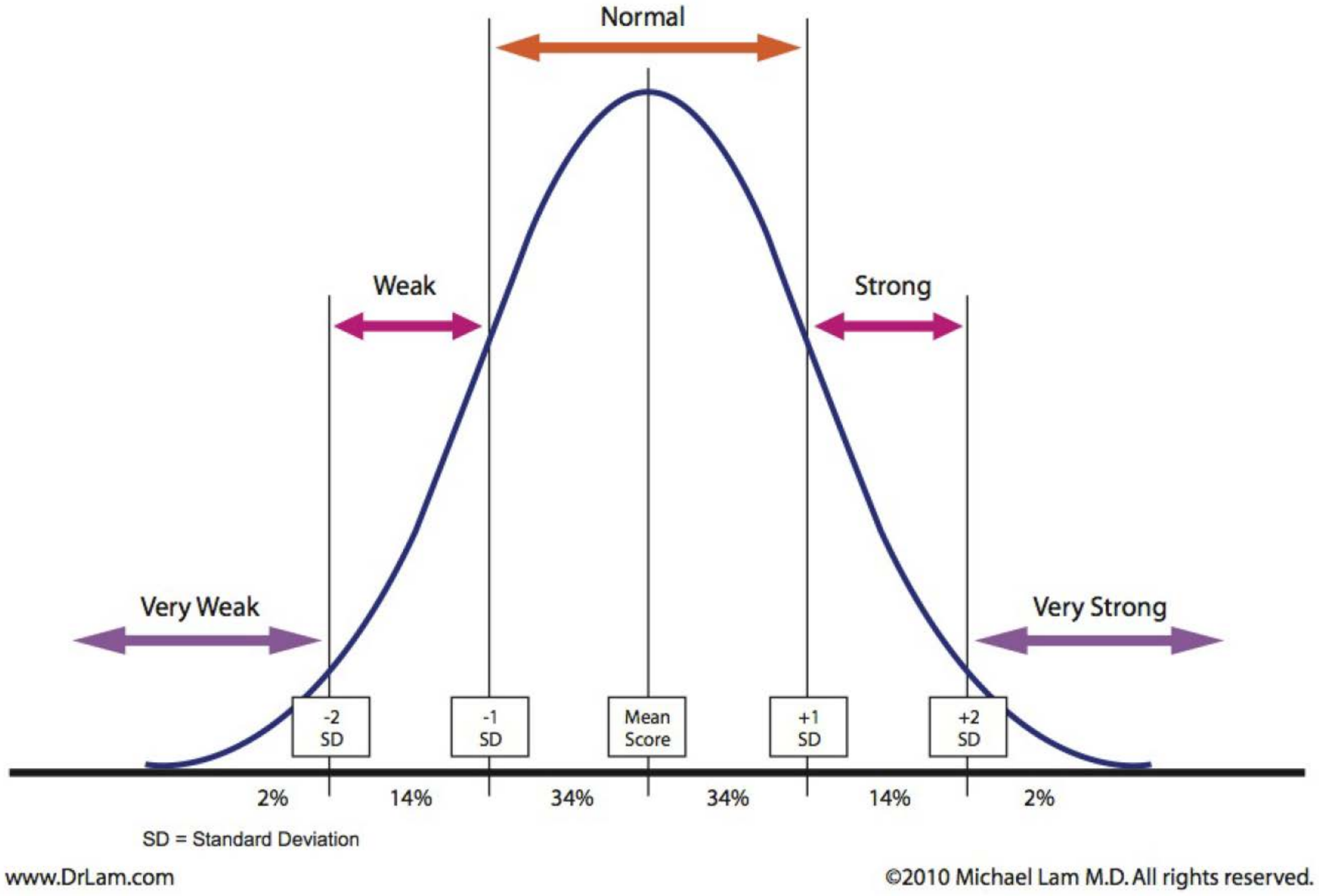}}
\caption{The normal probability curve which describes human IQ scores 
and many other randomly distributed quantities. The height of the curve 
for any score gives the relative probability for a randomly chosen human 
to achieve that score. The percentage figures on the $x$-axis give the
fraction of the total population whose scores fall within this range.}
\label{Normal}
\end{figure}

Educated readers are familiar with the concept of quantities which are 
distributed randomly such as human intelligence. Figure~\ref{Normal} 
shows the famous ``bell curve''. The probability that a randomly selected
human's IQ score $i$ lies between and two values $i_1$ and $i_2$ is given
by the area under the curve between $i_1$ and $i_2$. What quantum 
mechanics does is to make every conjugate pair, like $x_0$ and $p_0$ 
above, into random variables whose probability densities are inversely 
related so that when one variable is very well known, the other is poorly 
known. To understand precisely how this works requires a little integral 
calculus which is explained in Appendix A. The important thing for our
discussion is that, if the value of $x_0$ is known to an uncertainty of 
$\pm \Delta x$, and the value of $p_0$ is known to an uncertainty of $\pm 
\Delta p$, then the product of the two uncertainties obeys a famous 
inequality known as the Uncertainty Principle,
\begin{equation}
\Delta x \times \Delta p \geq \hbar \; . \label{uncert}
\end{equation}

The constant $\hbar \approx 1.054 \times 10^{-34}~{\rm kg\!\!-\!\!m}^2{
\rm /s}$ is very small when measured in terms of normal units, which is 
why we perceive no quantum effects in every day life. For example, most
people would consider that a measurement of some object's position was 
exceptionally accurate if it determined to within $\Delta x_0 = 
10^{-10}~{\rm m}$, but this would still allow the same object's momentum
to be measured to within $\Delta p_0 \simeq 10^{-24}~{\rm kg\!\!-\!\!m/s}$, 
which also seems wonderfully accurate. But the fact that we cannot make 
both $\Delta x_0$ and $\Delta p_0$ zero has terrific importance. One simple 
consequence is the stability of atoms against the collapse of their
electrons onto the central nucleus, which is energetically favoured
in classical physics. 

Consider the simplest atom, Hydrogen, whose energy is,
\begin{equation}
E = \frac{p^2}{2 m} - \frac{e^2}{4\pi \epsilon_0 \vert x\vert} \; .
\label{Hydrogen}
\end{equation}
In classical physics we can make the energy arbitrarily negative by
taking the electron-proton separation $x$ be close to zero {\it 
while keeping the kinetic energy smaller in magnitude.} This means it is 
energetically favorable for a classical Hydrogen atom to collapse and
emit a burst of electromagnetic radiation. The Uncertainty Principle is
what prevents that, because forcing $x$ to take any particular value
(such as zero) to within a small uncertainty $\Delta x$, causes the
uncertainty on the momentum to become large, which results in a very
large positive kinetic energy. This is easy to see from expression 
(\ref{Hydrogen}) if we make the energy as negative as possible by 
assuming that the average values of $x$ and $p$ are both zero, so that
only their uncertainties contribute,
\begin{equation}
E \longrightarrow \frac{\Delta p^2}{2 m} - \frac{e^2}{4 \pi \epsilon_0 
\Delta x}  \; .
\end{equation}
Now use the Uncertainty Principle to eliminate $\Delta p$ and make a
few simple algebraic rearrangements to derive a lower bound on the
energy,
\begin{equation}
E \geq \frac{\hbar^2}{2 m \Delta x^2} - \frac{e^2}{4 \pi \epsilon_0 
\Delta x} = \frac1{2 m} \Bigl( \frac{\hbar}{\Delta x} \!-\! \alpha m c
\Bigr)^2 - \frac12 \alpha^2 m c^2 \geq -\frac12 \alpha^2 m c^2 \; .
\label{bound} 
\end{equation}
The symbol $\alpha \equiv e^2/(4\pi \epsilon_0 \hbar c) \approx 1/137$
is known as the {\it fine structure constant} and we will encounter it 
again. The rest mass energy of the electron is usually expressed in
million electron-volts or MeV, $m c^2 \approx 0.511~{\rm MeV}$. Hence
our bound (\ref{bound}) implies that the energy of a {\it quantum
mechanical} Hydrogen atom cannot drop below the well known value,
\begin{equation}
E \geq -\frac12 \alpha^2 m c^2 \approx \frac12 \times \Bigl(\frac1{137}
\Bigr)^2 \times 511000~{\rm eV} \approx - 13.6~{\rm eV} \; .
\end{equation}
The agreement would not have been so perfect had we been more precise 
about defining what is meant by $\Delta x$, and about the atom living 
in three spatial dimensions, but none of this would change the fact
that the Uncertainty Principle stabilizes atoms against collapse.

\subsection{Quantum Field Theory}

A {\it field} is a dynamical variable which depends upon time $t$ 
as well as space $\vec{x}$. Familiar examples from introductory
physics are the electrodynamic scalar and vector potentials: 
$\Phi(t,\vec{x})$ and $\vec{A}(t,\vec{x})$. Classical fields have
initial value solutions, just like (\ref{qsol}), except that there 
are conjugate pairs for each space point $\vec{x}$. A {\it quantum 
field} obeys the same equation as its classical counterpart, the
only difference is that the conjugate pairs of its initial values
are random numbers which obey the Uncertainty Principle (\ref{uncert}).

The effect of quantization explains the otherwise puzzling statements
popular science readers may have seen that general relativity provides a 
classical theory of gravitation which agrees with all known observations, 
however, quantizing this perfect theory results in nonsense. That seems 
contradictory until one realizes that the wonderful classical agreement
comes from solutions for which {\it both} partners of almost all conjugate
pairs have been set to zero. In classical physics they are just numbers 
which can be fixed as we choose, but the Uncertainty Principle of quantum 
mechanics means there must be a minimum amount of disturbance in each 
conjugate pair. It is the cumulative effect of having each of the infinite 
number of conjugate pairs a little disturbed from zero which makes quantum 
general relativity so problematic. It is possible that we are not doing
the computation correctly (more on that in the next subsection) and that
nonlinear effects resolve the apparent difficulties. But it is easy to
understand how problems might arise from the addition of an infinite 
number of small contributions to the energy and momentum which sources 
gravitation.

We have a confession to make: {\it there is not any such thing as a
point particle}. In reality the wave functions of quantum mechanical
particles are just one linearized (and typically nonrelativistic limit 
of a) solution of a quantum field. So there is a quantum field which
describes each kind of particle: electrons, quarks, neutrinos, etc. A 
good way of categorizing these fields is by the spins of the particles 
they describe. Fundamental theory is subject to powerful constraints 
which limit the possibilities to just these:
\begin{itemize}
\item{{\it Spin 0 (scalar) fields} like the Higgs scalar which was 
recently discovered at the Large Hadron Collider;}
\item{{\it Spin $\frac12$ fermions} like the electron;}
\item{{\it Spin 1 (vector) fields} like electromagnetism or the weak
and strong nuclear forces;}
\item{{\it Spin $\frac32$ gravitinos} which occur in a conjectured
extension of general relativity known as supergravity; and}
\item{{\it Spin 2 (tensor) fields} of which the metric of general
relativity is the only possibility.}
\end{itemize}
Particles also have mass $m$, which is a characteristic of the type
of particle, as well as energy $E$ and momentum $\vec{p}$ which change
depending upon how the particle moves. An important relation (due to 
Einstein) exists between these three quantities in flat space (that
is, with $a(t) = 1$),
\begin{equation}
E = \sqrt{m^2 c^4 + \vec{p} \!\cdot\! \vec{p} \, c^2} \; . 
\label{Einstein}
\end{equation}
In the next section we will see how the Einstein relation (\ref{Einstein})
changes in an expanding universe. 

We reiterate that all particles are really excitations of quantum fields,
and hence probability waves. This has been tested past the point of any
dispute. For example, electrons have been made to show double slit 
interference patterns, just like light waves. Another important relation
(this one due to Louis DeBroglie) connects the wave length $\lambda$ to 
the magnitude $p$ of the momentum,
\begin{equation}
\lambda = \frac{\hbar}{p} = \frac{\hbar c}{\sqrt{E^2 \!-\! m^2 c^4}}
= \frac{\hbar}{\sqrt{2 m K \!+\! \frac{K^2}{c^2}}} \; . \label{DeBroglie}
\end{equation}
The final form on the right is expressed in terms of the kinetic energy 
$K \equiv E - m c^2$, and it explains why we don't normally recognize the 
wave nature of electrons or other particles with nonzero mass. Suppose the
electron has the amount of kinetic energy it is pretty much guaranteed to
have from thermal motion at room temperature, which is $K \approx 6 \times
10^{-21}~{\rm Joules}$. Substituting this into expression (\ref{DeBroglie})
gives a wave length of only $\lambda \approx 10^{-9}~{\rm m}$ which is
about 500 times smaller than visible light. We can directly perceive the 
wave nature of massless particles such as the photons which constitute 
electromagnetic radiation, and gravitons which constitute gravitational 
radiation. However, quantum mechanics offers a surprise for these radiation
fields in that they really consist of discrete quanta. According to 
relations (\ref{Einstein}-\ref{DeBroglie}) light or gravity waves of
wave length $\lambda$ corresponds to discrete particles of energy $E =
\hbar c/\lambda$. You can have this amount of energy in the radiation 
field, or any multiple of this amount, but you could not have 1.5 of this
energy, or $\sqrt{2}$ times this energy. This discrete nature of light
was demonstrated long ago in the photo-electric effect, which is what
powers solar cells. The quantization of gravitational radiation has not
yet been demonstrated, and the gravitational interaction is too weak for
individual gravitons to be seen at the LIGO detector of Figure~\ref{LIGO}. 
However, we will learn in the next section that cosmological measurements
can establish the quantization of gravitational radiation. With a bit of
luck, they might even establish the existence of {\it classical} 
gravitational radiation before LIGO detects it!

The spooky effects of quantum field theory all derive from the undoubted
fact that each conjugate pair in a {\it quantum} field is a little 
bit disturbed. This little bit of disturbance causes even empty space
to behave as if it was filled with classical particles, of all different
wave lengths $\lambda = \hbar/p$, which exist for a time $\Delta t$ 
governed by the {\it energy-time uncertainty principle}
\begin{equation}
\Delta t \leq \frac{\hbar}{E} = \frac{\hbar}{\sqrt{m^2 c^4 \!+\! 
\hbar^2 c^2 k^2}} \; , \label{ET} 
\end{equation}
where $k \equiv 2\pi/\lambda$ is called the wave number of the particle.
The particles which obey relation (\ref{ET}) are known as {\it virtual 
particles} and, once their existence is accepted, one can understand 
quantum field theory effects just by using classical physics. Of course 
that had to be true because quantum fields obey exactly the same 
equations as classical fields do; the only difference is that each of
their conjugate pairs must be a little nonzero, which is precisely what 
virtual particles represent.

\vskip -2cm
\begin{figure}[htp]
\centering{\includegraphics[scale = 0.7]{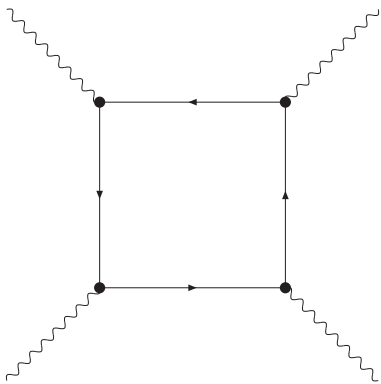}}
\vskip -13cm
\caption{Feynman diagram representing the lowest order process 
contributing to the scattering of two photons.}
\label{LbL}
\end{figure}
\vskip 0cm

As an example, consider the scattering of light by light. Many readers
learned in high school physics that two light rays pass right through one
another in empty space, with no effect at all. With virtual particles
that is no longer quite true. Figure \ref{LbL} is a {\it Feynman diagram}
which represents two quantized light particles (known as photons) coming 
in from the right. Then they are absorbed by a virtual electron and 
anti-electron (known as a positron) which exist for a brief period of
time given by relation (\ref{ET}). When the virtual pair's short lifetime 
is up, the two photons are re-emitted to the left, but with slightly 
different momenta than they originally had.

All quantum field theory effects can be understood in this way. And it
should be obvious that whatever strengthens the lifetime $\Delta t$ ---
and also the rate at which virtual particles are emitted --- will make
quantum field theory effects become stronger. We will see in the next 
section that the expansion of the universe does this. However, even in
flat space one can see from relation (\ref{ET}) that making the mass 
smaller increases $\Delta t$ at fixed $k$. That is why the strongest
quantum field theoretic effects come from the lightest particles. One
can also see that making the wave length long at fixed $m$ increases 
$\Delta t$.

\subsection{Perturbation Theory}

Physicists are very good at math, but most of the problems in general
relativity and quantum field theory are harder than we can solve. What 
we do instead is to develop approximate solutions in the form of power 
series expansions around the answers to simpler problems whose exact 
solution is known. Educated readers are familiar with the basic technique 
from computing square roots using the expansion,
\begin{equation}
\sqrt{1 + x} = 1 + \frac{x}2 - \frac{x^2}{8} + \frac{x^3}{16} - \dots
= 1 + \frac{x}{2} - \sum_{n=2}^{\infty} \frac{(2n \!-\! 3)!!}{n!} 
\Bigl( -\frac{x}{2}\Bigr)^n \; . \label{Taylor}
\end{equation}
(The double factorial symbol means to skip every other number, for
example, $7!! = 7 \cdot 5 \cdot 3 \cdot 1 = 105$.) For example, we know 
that the square root of 4 is exactly 2, and the square root of 9 is 
exactly 3, but we don't have exact results for the numbers in between. 
Suppose we want the square root of 5. We can get a series expansion by
first writing 5 as $4+1$, then factoring out the 4 and using expression 
(\ref{Taylor}) with $x = \frac14$,
\begin{eqnarray}
\sqrt{5} &=& \sqrt{4 + 1} = 2 \sqrt{1 + \frac14} = 2 + \frac14 - 2
\sum_{n=2}^{\infty} \frac{(2 n \!-\! 3)!!}{n!} \Bigl(-\frac18\Bigr)^n
\; , \label{sqrt5A} \\
& = & 2 + \frac14 - \frac1{64} + \frac1{512} - \frac{5}{16384} 
+ \frac{7}{131072} - \dots \label{sqrt5B}
\end{eqnarray}
With a scientific calculator one sees that $\sqrt{5} \approx 2.23607$.
This six digit accuracy is achieved by just the five terms shown in
(\ref{sqrt5B}).

Physicists who study quantum field theory employ a similar approximation 
technique known as {\it the loop expansion}. The first term is called 
{\it tree order}, the second term is called {\it one loop}, the third 
term is called {\it two loop}, and so on. It is sometimes difficult even 
getting the tree order result, and computing loop corrections becomes
prohibitively difficult very fast, so computations of real things are
never done to more than 4-5 orders. This disturbs some people but it is
well to recall that physicists can only avoid error by constantly 
checking their predictions against experiment and observation, and there
are important practical limitations on how accurately these measurements
can be performed. One of the best measured quantities is the gyromagnetic
ratio of the electron (the ratio of its actual magnetic dipole moment to
the classical prediction) which is accurately known to 12 digits. That
corresponds to 4-5 orders in the loop expansion, and there is simply no
point to pushing the theoretical prediction any farther until the
measurement can be done with greater accuracy.

Every quantum field theory has a {\it loop counting parameter} which
quantifies the loop expansion. For quantum electrodynamics (QED) this
parameter is $\alpha/2\pi \approx 1/861$, where $\alpha \equiv e^2/(4 \pi 
\epsilon_0 \hbar c)$ is the fine structure constant we saw in equation
(\ref{bound}). A typical prediction of QED takes the form,
\begin{equation}
\Bigl({\rm QED\ Prediction}\Bigr) = \Bigl({\rm Tree\ Result}
\Bigr) \Biggl\{1 + A \frac{\alpha}{2\pi} + B \Bigl(
\frac{\alpha}{2\pi}\Bigr)^2 + C \Bigl(\frac{\alpha}{2 \pi}\Bigr)^3 + 
\dots\Biggr\} \; .
\end{equation}
The ``Tree Result'' could also involve $\alpha$. The coefficients 
$A$, $B$ and $C$ can involve other parameters of the problem but they
are assumed to be of order one, so one can see that the 4-loop correction 
$(\alpha/2\pi)^4 \approx 1.8 \times 10^{-12}$ really does give 12-digit
accuracy, which is the best we can currently measure.

As one might expect, the loop counting parameter of quantum gravity
involves Newton's constant of universal gravitation, $G$. With the
other factors it takes the form $\hbar G \omega^2/c^5$, where $\omega$
is the characteristic frequency of whatever process is being studied.
For cosmological particle production that frequency is the Hubble
parameter, $H(t)$. From expression (\ref{H0}) one sees that the current
effect is hopelessly too small to resolve,
\begin{equation}
\frac{\hbar G H^2_0}{c^5} \approx 10^{-123} \; .
\end{equation}
However, the Hubble parameter $H_i$ of primordial inflation might be 
as much as 56 orders of magnitude larger. The best current upper bound 
is \cite{Planck:2015xua},
\begin{equation}
\frac{\hbar G H_i^2}{c^5} < 4 \times 10^{-11} \; . \label{infHub}
\end{equation} 
This number is interesting because it is says that quantum gravitational
effects are small during primordial inflation, but within our ability to 
measure. The fact that the number is small means we don't need to worry
about more than the first two terms.

\section{Primary Effects}\label{primary}

Quantum gravitational effects from primordial inflation are observable
for three reasons:
\begin{enumerate}
\item{The quantum gravitational loop counting parameter associated with
the rate of inflationary expansion (\ref{infHub}) is small enough that
we can compute reliably with quantum general relativity, but large enough
that sensitive measurements can resolve effects which involve one or two
powers of it;}
\item{Accelerated expansion allows significant numbers of virtual 
gravitons and the simplest type of massless scalars to emerge from
empty space and persist forever; and}
\item{The fluctuations these particles induce become fossilized so that
they can survive to be observed at late times.}
\end{enumerate}
We have already discussed the first point. In this section we will give
a simple way of understanding the second and third points. Then we discuss
the primordial power spectra which represent the primary effects.  

\subsection{Inflationary Particle Production}

We learned in section 2.4 that any particle is really a probability 
wave. It can have an arbitrary length $\lambda$, which fixes its energy 
and momentum by relations (\ref{Einstein}) and (\ref{DeBroglie}) in
flat space. In the cosmological geometry (\ref{FLRW}) a fixed wave 
length $\lambda$ corresponds to a physical length of $a(t) \times 
\lambda$, so it should not be surprising that the energy of a particle 
with wave length $\lambda = 2\pi/k$ is,
\begin{equation}
E(t,k) = \sqrt{m^2 c^4 \!+\! \Bigl(\frac{\hbar c k}{a(t)}\Bigr)^2 } 
\label{E(t,k)} \; .
\end{equation}
One interesting consequence is that the expansion of the universe 
lowers the energy of a particle.

Another thing we learned in section 2.4 is that quantum field 
theoretic effects can be understood as the classical response to 
virtual particles which emerge from empty space at time $t$ and
persist until time $t + \Delta t$. In flat space the persistence 
time $\Delta t$ is given by relation (\ref{ET}), but the alert reader 
will note a problem in that the cosmological energy (\ref{E(t,k)})
changes with time! What time should we evaluate it at? Stating the
answer precisely requires some integral calculus which is explained
in Appendix B. However, it should not seem surprising that the answer
involves a sort of average over all of the changing energies between
$t$ and $t + \Delta t$. 

Because $E(t,k)$ decreases as the universe expands, the persistence 
time $\Delta t$ is increased by cosmological expansion. As long as 
the particle's mass $m$ is nonzero its energy (\ref{E(t,k)}) cannot 
be smaller than $m c^2$, so $\Delta t$ remains finite. However, 
massless particles have $E(t,k) = \hbar c k/a(t)$, and it can be that 
the energy goes to zero so rapidly that the persistence time $\Delta t$ 
becomes infinite. In Appendix B we demonstrate that the condition for
this is inflation (which means negative deceleration parameter $q(t)$) 
with $c k < -q(t) H(t) a(t)$.

The combination $c/H(t)a(t)$ is known as the {\it Hubble radius}, and it 
represents roughly the physical distance that a local observer can see. 
So the wave lengths we are discussing are of cosmological scale. That is
very important because physicists know that there is something wrong
with our understanding of quantum general relativity at scales smaller
than the Planck length, $\sqrt{\hbar G/c^3} \sim 10^{-35}~{\rm m}$. Using
quantum general relativity at these scales or smaller would not be reliable.
On the other hand, using it at much larger scales should be right. (This
is yet another one of those correspondence limits which are so important
to physics!) Fortunately for us, the smallest Hubble radius for primordial 
inflation which is consistent with the observations of Figure \ref{Planck} 
is about $c/H_i a_i \sim 10^{-30}~{\rm m}$. That is very small, but still a
hundred thousand times larger than the Planck length, so we should be able
to trust quantum general relativity.

Just because long wavelength massless virtual particles persist forever
during inflation does not mean they necessary mediate significant quantum
field theoretic effects. One must still check that there is no suppression
of the {\it rate} at which these particles emerge from empty space. In 
flat space this rate is simply a constant but, as with many things, it 
can become time dependent in the cosmological geometry (\ref{FLRW}). A
simple way to analyze the problem is by changing the time coordinate from
$t$ to the {\it conformal time} $\eta$ which obeys the differential 
equation $dt = a(t) d\eta$. In terms of conformal time the cosmological
geometry (\ref{FLRW}) is the same as the flat space geometry (\ref{flat})
up to an overall multiplicative constant,
\begin{equation}
-c^2 dt^2 + a^2(t) d\vec{x} \!\cdot\! d\vec{x} = a^2 \Bigl[ -c^2 d\eta^2
+ d\vec{x} \!\cdot\! d\vec{x}\Bigr] \; .
\end{equation}   
Almost all massless particles possess a symmetry known as {\it conformal
invariance} which states that they are the same in conformal coordinates
as in flat space. This means that the rate at which conformally invariant
particles emerge from empty space per unit conformal time is the same
constant $\Gamma_{\rm flat}$ as it is in flat space. It follows that the
emission rate per unit physical time falls off with $a(t)$,
\begin{equation}
\frac{dN}{dt} = \frac{d\eta}{dt} \!\times\! \frac{dN}{d\eta} = 
\frac1{a(t)} \!\times\! \Gamma_{\rm flat} \; . \label{conformal}
\end{equation}
Therefore, any long wavelength conformally invariant virtual particle 
which emerges from empty space during inflation can persist forever, but 
very few emerge. There are just two sorts of massless particles which 
are not conformally invariant, and so avoid the reduced emission rate of 
expression (\ref{conformal}). They are the simplest kind of spin zero 
particles, and the particles associated with quantized gravitational 
radiation, which are known as gravitons.

\subsection{How Inflationary Perturbations Fossilize}

Both gravitons and massless minimally coupled scalars of wave number $k$
have the same wave function $u(t,k)$. It obeys the same equation as a 
point mass attached to the same sort of spring whose force law $F = -m 
\omega^2 x$ is familiar from introductory physics. The details are 
explained in Appendix C but the key point is that the mass and 
characteristic frequencies of this oscillator are time dependent,
\begin{equation}
m(t) = \sqrt{\frac{32\pi G \hbar}{c}} \times a^3(t) \qquad \qquad
\omega(t,k) = \frac{c k}{a(t)} \; . \label{SHO}
\end{equation}
For inflation the frequency $\omega(t,k)$ falls off so rapidly that the
particle stops oscillating because the force acting on it gets so small. 
However, the fact that the mass $m(t)$ grows more rapidly that 
$\omega^2(t,k)$ falls off means that the system freezes in with a
substantial amount of potential energy. One can count the number $N(t,k)$ 
of particles produced by simply considering this potential energy to be
$N(t,k)$ times the energy $\hbar c k/a(t)$ of a single particle.

Explicit computations are not possible for general $q(t)$ but Figure
\ref{decel} shows it should be a good approximation to set $q_i = -1$, 
which means $H(t)$ is some constant $H_i$. These approximations reveal 
a staggering amount of particle production,
\begin{equation}
q_i = -1 \qquad \Longrightarrow \qquad N(t,k) = \Bigl[ \frac{H_i a(t)}{2 c k}
\Bigr]^2 \; . \label{dSN}
\end{equation}
Note that the particle number is small initially when $ck \gg H(t) a(t)$, 
it becomes of order one at $t_k$, the time of {\it first horizon crossing} 
when $c k = H(t_k) a(t_k)$, and it grows explosively for late times.
Under the same approximation one can also show that the amplitude of
$u(t,k)$ approaches the constant,
\begin{equation}
q_i = -1 \qquad \Longrightarrow \qquad \lim_{t \rightarrow \infty} \vert
u(t,k) \vert = \frac{H_i}{c k} \, \Biggl[\frac{8 \pi \hbar G}{c^3 k^2}
\Biggr]^{\frac14} \label{fossil}
\end{equation}
The fact that $u(t,k)$ approaches a constant is how perturbations from
gravitons and the simplest kind of massless scalars become fossilized so 
that they can be seen at much later times.

The process of fossilization does not depend upon the assumption we made
in expressions (\ref{dSN}-\ref{fossil}) that $q(t) = -1$. To see this, 
consider the time derivative of the product of $H(t)$ and $a(t)$ for 
general $q(t)$,
\begin{equation}
\frac{d}{dt} \Bigl[ H(t) a(t)\Bigr] = \frac{d}{dt} \Bigl[ \dot{a}(t)\Bigr]
= \ddot{a}(t) = -\Bigl( -\frac{a \ddot{a}}{\dot{a}^2}\Bigr) 
\Bigl(\frac{\dot{a}}{a}\Bigr)^2 a = -q(t) H^2(t) a(t) \; . \label{Hubrad}
\end{equation}
From expression (\ref{Hubrad}) we see that $H(t) a(t)$ increases during
inflation ($q(t) < 0$) and decreases during deceleration ($q(t) > 0$). 
The wave numbers $k = \frac{2\pi}{\lambda}$ of interest for us pass through
three stages of evolution over the course of the cosmological history 
depicted in Figure \ref{decel}:
\begin{enumerate}
\item{At the beginning of primordial inflation they obey $c k \gg H(t) a(t)$.
While this is true the wave function $u(t,k)$ oscillates rapidly and its 
magnitude falls off like $1/a(t)$. During this phase the occupation number 
$N(t,k)$ is very small.}
\item{First horizon crossing $ck = H(t_k) a(t_k)$ occurs at some time $t_k$
about 50 e-foldings before the end of primordial inflation. After this point
the characteristic frequency $\omega(t,k) = c k/a(t)$ becomes negligible 
with respect to the Hubble parameter $H(t)$ and the wave function $u(t,k)$ 
approaches a constant. After first horizon crossing the occupation number 
$N(t,k)$ grows rapidly.} 
\item{Second horizon crossing $ck = H(T_k) a(T_k)$ occurs at some time 
$T_k$ after the end of primordial inflation and before the onset of the
current phase of cosmic acceleration. At this point the wave function
$u(t,k)$ begins to change with time again, although according to the
altered conditions which prevail at late times.}
\end{enumerate}
Of course there are large $k$ wave numbers which never experienced first 
horizon crossing. We can measure them at current times but they are not very 
interesting because they did not become highly populated during primordial
inflation, nor did they ever fossilize. There are also some small $k$ wave
numbers which have passed through first horizon crossing but never experienced
second horizon crossing. These wave numbers are still fossilized and their 
physical wave lengths are so large that the limited portion of the universe
we can see does not allow us to perceive their spatial variation. People who
think they know things without doing calculations or measurements claim that 
these wave numbers can have no effect at all today but it seems safer to
conclude that whatever effect these modes have must seem spatially (but 
perhaps not temporally) constant to us. There may also be some very small 
$k$ wave numbers which were fossilized when primordial inflation began and 
are still fossilized today.

\subsection{The Primordial Power Spectra}

There are two primary effects: the ensemble of gravitons which are produced 
during inflation and the gravitational response to inflation-enhanced quantum 
fluctuations in whatever matter fields provided the source for primordial 
inflation. The first effect is weaker than the second but conceptually 
simpler. It was originally described by the brilliant Russian cosmologist 
Alexei Starobinsky in 1979 \cite{Starobinsky:1979ty}. This effect is usually 
represented by a quantity known as the {\it tensor power spectrum} 
$\Delta^2_{h}(k)$ which can be expressed in terms of the wave function 
$u(t,k)$,
\begin{eqnarray}
\Delta^2_{h}(k) & = & \frac{k^3}{2 \pi^2} \!\times\! 2 \!\times\!
\sqrt{\frac{32 \pi \hbar G}{c^3}} \times\! \Bigl\vert u(t,k)\Bigr\vert^2_{t_k 
\ll t \ll T_k} \Biggl\{1 + \Bigl({\rm Loop \atop Corrections}\Bigr)\Biggr\} 
, \qquad \label{exacttens} \\
& \approx & \frac{16 \hbar G H^2(t_k)}{\pi c^5} \Biggl\{ 1 + \# 
\frac{\hbar G H^2}{c^5} + \dots \Biggr\} . \label{approxtens}
\end{eqnarray}
Because $u(t,k)$ in expression (\ref{exacttens}) is evaluated long after
first horizon crossing and long before second horizon crossing, the wave
function is at its constant, fossilized value. The approximation
(\ref{approxtens}) was obtained by substituting the $q(t) = -1$ result
(\ref{fossil}) into expression (\ref{exacttens}). At this stage no one is
sure how to compute loop corrections (debates about this sometimes degenerate
into shouting matches!) although these corrections should be suppressed by 
the quantum gravitational loop counting parameter (\ref{infHub}). One of the 
many things which is not clear is {\it when} the slightly time dependent 
Hubble parameter should be evaluated.

The gravitational response to quantum fluctuations in whatever matter drove
primordial inflation is usually quantified as the {\it scalar power spectrum} 
$\Delta^2_{\mathcal{R}}(k)$. The original computation of it was made by
Viatcheslav Mukhanov (now at Ludwig Maximilian University in Munich, Germany) 
and Gennady Chibisov (now deceased) \cite{Mukhanov:1981xt}. Of course the result 
for it depends upon precisely {\it what} matter fields caused primordial 
inflation, which is a subject of intense debate. In the simplest model 
$\Delta^2_{\mathcal{R}}(k)$ depends upon a wave function we might call 
$v(t,k)$ which behaves as an oscillator with the same characteristic frequency 
(\ref{SHO}) as $u(t,k)$, but a much smaller mass $M(t) = \sqrt{\frac{c \hbar}{
4 \pi G}} [1 + q(t)] a^3(t)$. This oscillator is much lighter than $u(t,k)$ 
because $q(t)$ was so near $-1$ during primordial inflation. Of course being 
lighter means the amplitude is larger! In terms of $v(t,k)$ the scalar power 
spectrum takes the form,
\begin{eqnarray}
\Delta^2_{\mathcal{R}}(k) & = & \frac{k^3}{2 \pi^2} \!\times\!
\sqrt{\frac{4 \pi \hbar G}{c^3}} \times\! \Bigl\vert v(t,k)\Bigr\vert^2_{t_k 
\ll t \ll T_k} \Biggl\{1 + \Bigl({\rm Loop \atop Corrections}\Bigr)\Biggr\} 
, \qquad \label{exactscal} \\
& \approx & \frac{\hbar G H^2(t_k)}{\pi c^5 [1 \!+\! q(t_k)]} \Biggl\{ 1 + \# 
\frac{\hbar G H^2}{c^5} + \dots \Biggr\} . \label{approxscal}
\end{eqnarray}
There are the same passionate debates about how to calculate loop corrections
to $\Delta^2_{\mathcal{R}}(k)$, and the same uncertainty about what time to
evaluate the factors of $H(t)$ which appear in the loop counting parameter.

The scalar power spectrum was first resolved in 1992 by the Cosmic Background
Explorer (COBE) satellite \cite{Smoot:1992td}, for which George Smoot and John 
Mather shared the 2006 Nobel Prize. This first detection measured only a small 
range of the largest interesting wavelengths $\lambda = 2\pi/k$. Since then 
there have been two major satellite probes and scores of balloon-bourne detectors 
which have resolved wavelengths more than a thousand times smaller than those
detected by COBE. The result is well fit by a function of the form 
\cite{Planck:2015xua},
\begin{equation}
\Delta^2_{\mathcal{R}}(k) = A_s \Bigl( \frac{k}{k_0}\Bigr)^{n_s-1} \; ,
\end{equation}
where the amplitude $A_s$ and scalar spectral index $n_s$ are,
\begin{equation}
A_s = \Bigl(2.142 \pm 0.049\Bigr) \!\times\! 10^{-9} \qquad , \qquad 
n_s = 0.9667\pm 0.0040 \; .
\end{equation}
The pivot $k_0 = 0.050~{\rm Mpc}^{-1}$ corresponds to a wavelength of
about $\lambda_0 = 3.9 \times 10^{24}~{\rm m}$. The fact that the scalar
spectral index $n_s$ is so close to one means that different wavelengths
have almost the same amplitude. By comparison with the theoretical prediction
expression (\ref{approxscal}) we can infer that the inflationary Hubble 
parameter $H(t)$ and deceleration parameter $q(t)$ were nearly constant 
during the approximately ten e-foldings over which the measured wavelengths
experienced first horizon crossing. This is thought to have occurred about
50 e-foldings before the end of primordial inflation.

\begin{figure}[htp]
\centering{\includegraphics[angle= -90,scale = 0.40]{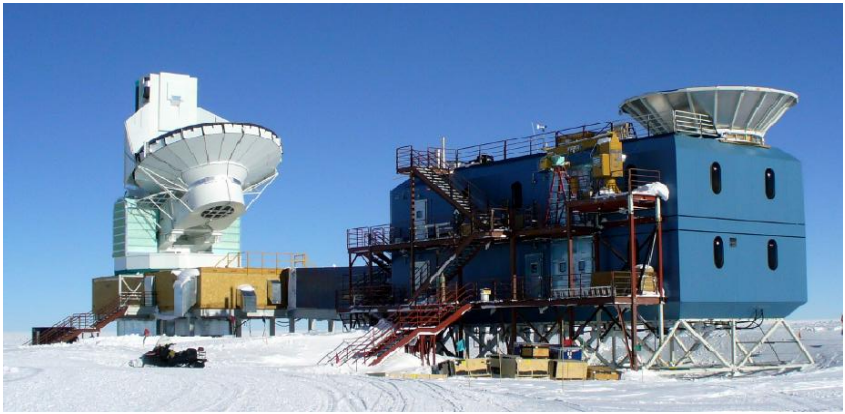}}
\caption{The BICEP2 detector located at the South Pole.}
\label{bicep2}
\end{figure}

The tensor power spectrum has not yet been resolved. This is presumably
due to the fact that the inflationary deceleration factor $q(t)$ is so 
close to $-1$, which makes the scalar power spectrum much larger than the 
tensor power spectrum. In March of 2014 the people who run the BICEP2 
detector (depicted in Figure \ref{bicep2}) announced that they had 
succeeded in resolving $\Delta^2_{h}(k)$ with huge statistical 
significance \cite{Ade:2014xna}. However, it turns out that they were 
mostly detecting the light from dust near our galaxy 
\cite{Adam:2014bub,Ade:2015tva}. The team did under-estimate the 
contamination from dust, and they certainly over-publicised their 
findings, but it's hard to fault them much more because they did 
manage to achieve a huge increase in sensitivity, and the people who 
possessed the best dust map (paid for by European taxpayers) had refused 
to make it public. When the dust map was finally released and it became
apparent that their signal could be entirely due to dust, the BICEP2
team admitted their mistake and actually collaborated with their critics
to produce the best current upper limit for the strength of the tensor
power spectrum. The whole exercise has been a wonderful stimulant
for making even more sensitive measurements in regions which are now
known to be relatively free of dust.

No one knows when --- or if --- the tensor power spectrum will be 
resolved but there are many ground-based and balloon-bourne detectors 
which will report data over the next few years, and hopes are high.
Resolving the tensor power spectrum is crucially important because it
will tell us the scale of primordial inflation. Resolving $\Delta^2_h(k)$
over many wavelengths will also allow us to test an entire class of
models for the matter theory which caused primordial inflation.

\section{Secondary Effects}\label{secondary}

As we saw from equation (\ref{dSN}), accelerated expansion leads to
the explosive production of the simplest kind of massless scalars
and of gravitons. This is what causes the scalar and tensor power
spectra (\ref{exactscal}) and (\ref{exacttens}). However, it is just 
about inconceivable that these particles do nothing after having 
been produced. Normal particles interact with one another, and with 
other types of particles. Indeed, a particle could never be detected 
if it failed to interact with anything.

We call these interactions {\it secondary effects}. Because gravity 
is a weak interaction --- even at the scales of primordial inflation
--- scalar secondary effects are typically stronger than those from 
gravitons. However, graviton effects are more generic because there 
are many, many different scalar theories --- including the possibility 
there are none of the simplest kind of scalars --- whereas there is 
only one theory of gravity and we know it applies. 

Quantum field theory is a very tough subject, even for people who 
spend their lives studying physics. Some of the work described in this
section required derivations of over 80 pages in the highly condensed 
language of theoretical physics! We shall therefore confine the 
discussion of this section to the barest sketch of what was found,
why it happens and what it means, but we will cite the primary literature
and mention the nationalities of the authors. For each class of secondary 
effect (those caused by scalars and those caused by gravitons) we will 
follow the hierarchy of listing secondary effects on scalars, fermions, 
vector particles and gravitons.   

\subsection{Effects Mediated by Scalars}

The best studied model is when the simplest kind of scalar field is
given the simplest kind of interaction and studied without gravitational
interactions in the simplest inflationary geometry. For this {\it
simple} model the inflationary production of scalars increases the
scalar field strength. That causes the following secondary effects:
\begin{itemize}
\item{The energy density of the scalar field increases, just as it
would classically if the scalar field strength were increased
\cite{Onemli:2002hr,Onemli:2004mb}.}
\item{Scalar particles develop an increasing mass, just as they would 
classically when the scalar field strength increases 
\cite{Brunier:2004sb,Kahya:2006hc}.}
\end{itemize}
This model is the only one in which people have worked out the lowest
order quantum corrections to the initial state \cite{Kahya:2009sz} so
that the system can be released at a finite time and allowed to evolve. 
Physicists who have studied this model include the Frenchman Tristan 
Brunier, Turks Emre Kahya and Vakif Onemli, and American Richard Woodard.

No one knows whether or not the newly-discovered Higgs scalar is the
simplest kind of scalar, or some more complicated kind. Assuming it
is the simplest kind, then it will experience explosive particle
production during inflation. The effect on known fermions such as the 
electron has been studied. What happens is that the increasing scalar 
field strength causes the fermion to behave as if it had a nonzero mass 
--- just as a classical increase in the scalar would
\cite{Prokopec:2003qd,Garbrecht:2006jm}. That causes the 0-point energy
of each wave number of the fermion to become ever more negative
\cite{Miao:2006pn}. If nothing else counteracts this effect then the 
universe would blow up in a Big Rip singularity. Meanwhile, there is no 
large change in the scalar mass\cite{Duffy:2005ue}, but the way the 
scalar contributes to the energy density of the universe becomes far 
more complicated and interesting than in flat space \cite{Miao:2006pn}. 
Cited authors include the Australian Leanne Duffy, German Bjorn Garbrect, 
Taiwanese Shun-Pei Miao, Croatian Tomislav Prokopec, and American Richard 
Woodard.

The simplest kind of scalar might be charged. If one couples it to 
electrodynamics in the usual way then the increase of the scalar field 
strength causes the photon to develop an increasing mass --- just as a 
classical increase would \cite{Prokopec:2002jn,Prokopec:2002uw,
Prokopec:2003bx,Prokopec:2003iu}. While this happens the scalar remains 
light \cite{Prokopec:2006ue,Prokopec:2007ak}. However, the contribution 
that the entire system makes to the energy density of the universe {\it 
decreases} \cite{Prokopec:2007ak,Prokopec:2008gw}, for the same reason 
as a slab of dielectric is pulled inside a charged parallel plate 
capacitor in classical physics. One can also study how the ensemble of 
charged scalars produced by inflation modifies electric and magnetic 
forces. One dramatic effect is a rapid extinction of the force between 
two charges which are initially at rest with respect to one another 
\cite{Degueldre:2013hba}. This phenomenon of {\it charge screening} 
is exactly what would occur classically if there were some classical 
process to dump an ever-larger number of balanced, positive-negative
charge pairs into the region between the two charges. Researchers 
include the Swiss Henri Degueldre, Croatian Tomislav Prokopec, Swede 
Ola Tornkvist, Greek Nikolaos Tsamis and American Richard Woodard.

Finally we come to the effects of scalars on gravitons and on the 
force of gravity. Although inflation certainly produces an ensemble
of the simplest kind of scalars, direct computation show that they
have no significant effect on the propagation of gravitons
\cite{Park:2011ww,Park:2011kg,Leonard:2014zua}. The reason seems to 
be that gravitons interact with the {\it derivative} of the scalar 
field strength, and though the scalar field strength builds up, its 
derivative does not. If that reasoning is correct then a more 
complicated scalar might produce some effect. For example, if the
scalar had a very small mass it would still be produced efficiently
during inflation, but its field strength would matter. Study of a
related possibility is under way \cite{Boran:2014xpa}. We have 
cited work done by the Turks Sibel Boran and Emre Kahya, Americans
Katie Leonard and Richard Woodard, Korean Sohyun Park and 
Croatian Tomislav Prokopec.

What has not yet been worked out is the effect that scalars have
on the force of gravity. There must be some effect of this type
during inflation because there is one in flat space. (Another of
those wonderful Correspondence Limits which physicists find so 
useful.) The flat space effect takes the form of a fractional 
correction to the force between two masses $m_1$ and $m_2$ which 
are separated by a distance $r$,
\begin{equation}
F_{\rm flat}(r) = \frac{G m_1 m_2}{r^2} \Biggl\{ 1 + \# 
\frac{\hbar G}{c^3 r^2} + \dots \Biggr\} \; . 
\end{equation}
The physical interpretation is that virtual scalars tend to collapse
onto the sources, which makes them become stronger. Because more and
more scalars are produced during $q(t) = -1$ inflation there is the 
possibility for a growing correction of the form,
\begin{equation}
F_{\rm inf}(r) = \frac{G m_1 m_2}{r^2} \Biggl\{ 1 + \# 
\frac{\hbar G H^2_i}{c^5} \times H_i t + \dots \Biggr\} \; . 
\end{equation}
The formalism needed to check for such an effect has been derived
\cite{Leonard:2014zua} but it has not yet been used to check for the
effect.

\subsection{Effects Mediated by Gravitons}

Gravitons can affect themselves and other particles by scattering 
with them, much like the virtual electron-positron alter the motions
of two photons in Figure \ref{LbL}. As the external particle 
propagates further and further through the ensemble of inflationary 
gravitons, one would expect it to be affected more and more {\it 
provided} it continues to interact effectively with inflationary
gravitons. From expression (\ref{dSN}) we see that inflationary
gravitons typically have physical wave number $H_i/c$. However, the 
physical wave number of the external particle redshifts like $k/a(t)$. 
So finding a significant effect requires that the external particle 
must have some interaction with gravity in addition to its 
red-shifting energy $\hbar c k/a(t)$.

The simplest kind of scalars interact with gravity only through their
red-shifting energies, so they experience no significant effect
\cite{Kahya:2007bc,Kahya:2007cm}. It is possible that more complicated
scalars will show an effect, although this has not been checked. Work
on this subject has been done by Turkish citizen Emre Kahya and American
Richard Woodard.

Fermions possess a spin-spin interaction in addition to their kinetic 
energy, and this does not red-shift. One result is that fermions are 
gradually excited by inflationary gravitons until their field strengths 
become so large that perturbation theory is no longer valid 
\cite{Miao:2005am,Miao:2006gj,Miao:2007az,Miao:2008sp,Miao:2012bj}. It 
would be very interesting to see how this affects the 0-point energy 
density contributed by fermions. This system has been studied by
Taiwanese citizen Shun-Pei Miao and American Richard Woodard.

Photons also possess a spin-spin interaction with gravitons so one
would expect to find similar enhancements. This is indeed the case
when detailed computations are done to quantum-correct the Maxwell 
equations which readers learn in beginning physics \cite{Leonard:2013xsa}. 
Inflationary gravitons make fascinating changes in electric and 
magnetic forces because they add momentum to the virtual photons 
which carry these forces \cite{Glavan:2013jca}. For example, long 
range virtual photons carry only a tiny momentum, so the amount they 
can acquire from inflationary gravitons is large. Hence the long range 
electric field is strengthened \cite{Glavan:2013jca}. The same effect 
strengthens the electric fields of dynamical photons as they propagate 
during inflation \cite{Wang:2014tza}. A study is under way of how this
affects  the 0-point energy density contributed by photons. Work on
this system has been done by Croatians Drazen Glavan and Tomislav
Prokopec, Americans Katie Leonard and Richard Woodard, Taiwanese 
Shun-Pei Miao and Chinese citizen Changlong Wang.

Of course gravitons can interact with other gravitons through their
spin-spin couplings. The complexity of dealing with quantum gravity
has so far prevented anyone from quantum-correcting the equations of
gravity waves, but a simple approximation to these equations can be
made which gives qualitatively correct results in all cases for which 
it can be checked. When this approximation is made for dynamical
gravitons one finds that they too are excited by interaction with
inflationary gravitons \cite{Mora:2013ypa}. No result has yet been
obtained for what inflationary gravitons do to the force of gravity,
or how they affect the expansion of spacetime. The paper we have 
mentioned was co-authored by Americans Pedro Mora and Richard 
Woodard, and Greek citizen Nikolaos Tsamis.

Finally, we should note that these graviton effects are 
controversial, even though they seem to have sensible explanations.
There are reasons for this. For example, part of the gravitational
dynamical variable is arbitrary and it is sometimes difficult to be
certain this arbitrary part has not contaminated a calculation to
produce an incorrect result known as an {\it gauge artefact}. Some
people also worry that quantum field theory computations average
over portions of the wave function which have {\it decohered}, much
like the famous feline of the Schrodinger Cat Paradox. In that
thought-experiment one encloses a cat inside a box with a cannister
of poison gas which is triggered by a completely random, nuclear 
decay process. If the decay occurs before the box is re-opened the
gas is released and the cat dies. Averaging over a wave function 
which contains both the living and dead cat is a mistake because any
observer who checks the box will either see a live (and pretty angry)
cat or a dead cat, but not the average of both.

However, these worries can be checked, albeit with effort, and the 
checks never seem to satisfy the critics. The authors of this article
sometimes experience a feeling of dizziness when listening to their 
own careful calculations being dismissed as nonsense by the very same 
people who consider wild and untestable speculation with the greatest
solemnity. It is difficult to avoid the feeling that some researchers 
are so deeply and loudly suspicious of secondary effects for no other 
reason than that they find the idea unattractive. The stubborn 
resistance to black holes noted in the Introduction is more typical 
than exceptional in the history of science. Far too many physicists 
allow pre-conceived notions to color their scientific judgements, 
which is why we ask the public to insist that scientists back up 
their opinions with observation and experiment, or with careful 
reasoning using principles established by observation and evidence.

\begin{figure}[htp]
\centering{\includegraphics[scale = 0.20]{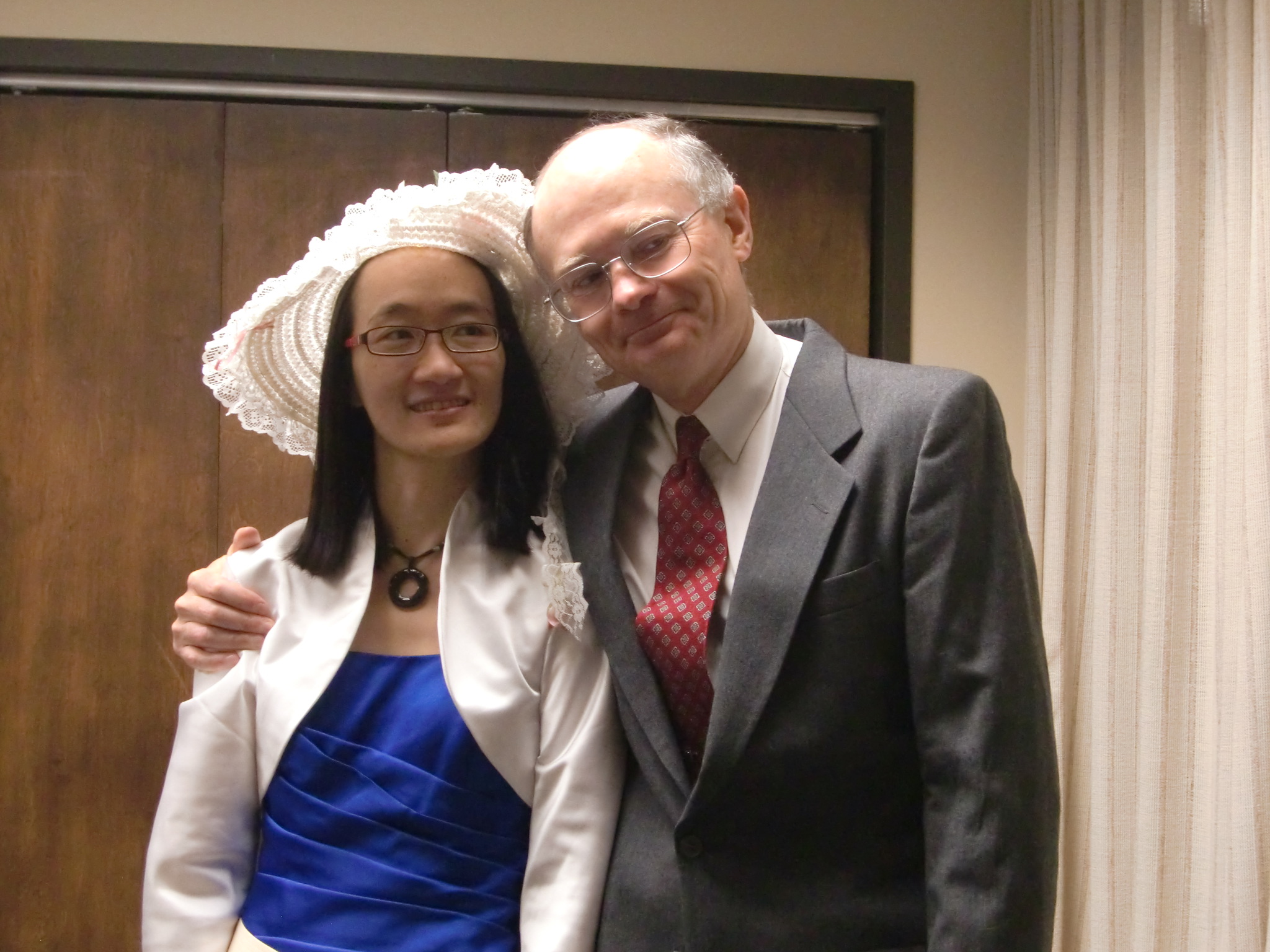}}
\caption{The authors of this article are a Taiwanese-American 
couple who began working together in 2004 and came to enjoy each
other's company so much that they married in 2013.}
\label{us}
\end{figure}

But let us not leave the impression of bitter frustration. If we
are right, the ever-more-precise and more abundant data will bear us 
out. In the meantime, vigorous criticism is one way science gradually
works its way towards truth, and the motives of critics ought not to
matter as long as the debate they spark is open. The authors of this 
article have been privileged to work on some fascinating problems at
public expense. And the occasional critic has not prevented us from 
enjoying life, as witness Figure \ref{us}.

\section{Conclusions}\label{conclusions}

Quantum gravitational effects from the epoch of primordial
inflation are observable for three reasons:
\begin{enumerate}
\item{The inflationary Hubble parameter (\ref{infHub}) is small
enough for perturbation theory to be wonderfully accurate, but not
so small that effects are beyond our ability to measure;}
\item{The accelerated expansion of primordial inflation produces
huge numbers (\ref{dSN}) of long wavelength gravitons and the
simplest type of massless scalars; and}
\item{The wave functions (\ref{fossil}) of these particles fossilize
during primordial inflation so that they can be preserved to be
observed at much later times.}
\end{enumerate}
These primary effects are quantified by the tensor and scalar power
spectra (\ref{exacttens}) and (\ref{exactscal}). The scalar power
spectrum has been resolved to 3-digit accuracy over wavelengths that
differ by a factor of about a thousand. The tensor power spectrum
has not yet been resolved but hopes are high that it soon will be.

Unless physics is radically different from what we believe, the
primary effects must give rise to much weaker secondary effects in
which inflationary scalars and gravitons alter the spacetime 
geometry and also change the dynamics of themselves and other 
particles. These secondary effects are too small to be measured now
but they may become observable in several decades when the data 
available in 21-cm radiation has been fully analyzed.

Deriving these secondary effects has been the work of many men and 
women, from nations all around the world, over the course of more 
than a decade. And the most interesting computations remain to be 
done! These are the effect of gravitons on the force of gravity and 
on the overall geometry.

Our ability to for the first time observe quantum gravitational
effects is changing fundamental theory. It has also confronted us
with three great open problems:
\begin{itemize}
\item{What caused primordial inflation?}
\item{What is causing the current phase of inflation? and}
\item{How do we correctly relate theoretical computations to 
observations?}
\end{itemize}
Finding the answer to any of these problems would merit a Nobel 
Prize. We close with the hope that someone in the vast population 
of Mandarin readers who might see this article will rise to the
challenge.

\vskip 1cm

\centerline{\bf Acknowledgements}

We are grateful for conversation and correspondence on this subject
with R. H. Brandenberger, S. Deser, L. H. Ford, W. Israel, E. O. Kahya, 
K. E. Leonard, P. J. Mora, S. Park, T. Prokopec, A. Roura, N. C. Tsamis,
C. L. Wang, S. Weinberg and H. L. Yu. This work was partially supported 
by Taiwan MOST grant 103-2112-M-006-001-MY3, by the Focus Group on 
Gravitation of the National Center for Theoretical Sciences of Taiwan, 
by NSF grants PHY-11-25915 and PHY-1205591, and by the Institute for 
Fundamental Theory at the University of Florida.

\section{Appendix A: Quantum Mechanical Probability Densities}

Figure~\ref{Normal} shows the famous ``bell curve'' $\rho(i)$ whose 
integral gives the probability that a randomly selected human's IQ score 
$i$ will lie between any two values $i_1$ and $i_2$,
\begin{equation}
{\rm Prob}\Bigl( i_1 \leq i \leq i_2\Bigr) = \int_{i1}^{i_2} \!\!\!\!
di \, \rho(i) \; .
\end{equation}
We can be precise about the relation between the probability densities
of conjugate pairs for readers who understand complex numbers and integral 
calculus. The state of the quantized system (\ref{qsol}) is described by a 
single complex-valued wave function $\psi(x_0)$ which is normalized,
\begin{equation}
\int_{-\infty}^{\infty} \!\!\!\!\! dx_0 \, \Bigl\vert \psi(x_0) 
\Bigr\vert^2 = 1 \; . \label{normal}
\end{equation}
As long as it obeys condition (\ref{normal}), the wave function can be
anything. However, once the wave function is known, it fixes the probability
densities for both members of a conjugate pair as follows:
\begin{eqnarray}
\rho(x_0) & \equiv & \Bigl\vert \psi(x_0) \Bigr\vert^2 \; , 
\label{pos} \\ 
\widetilde{\rho}(p_0) & \equiv & \Biggl\vert \int_{-\infty}^{\infty} 
\!\!\!\!\! dx_0 \, \exp\Bigl[\frac{i x_0 p_0}{\hbar} \Bigr] \psi(x_0) 
\Biggr\vert^2 \; . \label{mom}
\end{eqnarray}
One can use relations (\ref{pos}-\ref{mom}) to prove the Uncertainty 
Principle (\ref{uncert}).

\section{Appendix B: Energy-time Uncertainty Principle in Cosmology}

In an expanding universe the energy (\ref{E(t,k)}) is no longer a 
constant so the energy-time uncertainty principle becomes an integral,
\begin{equation}
\int_{t}^{t+\Delta t} \!\!\!\!\! dt' E(t',k) \leq \hbar \; . 
\label{cosmoET} 
\end{equation}
The way to read expression (\ref{cosmoET}) is that a virtual particle
of wave number $k$, which emerges from empty space at time $t$, can
persist for a time $\Delta t$ given by the inequality. When $a(t) = 1$ 
the energy is constant so the integral gives just $\Delta t \times E$,
as in relation (\ref{ET}). Checking these sorts of correspondence
limits is one important way that physicists avoid making mistakes!

As long as the mass $m$ in expression (\ref{cosmoET}) is nonzero the
integral grows with $\Delta t$ (it is always more than $m c^2 \Delta t$)
and must eventually violate the inequality. So massive virtual particles 
live longer in an expanding universe, but they must eventually disappear.
However, the case of massless particles is interesting,
\begin{equation}
\lim_{m \rightarrow 0} \int_{t}^{t+\Delta t} \!\!\!\!\! dt' E(t',k) 
= \hbar c k \int_{t}^{t + \Delta t} \frac{dt'}{a(t')} \leq \hbar \; .
\label{masslessET}
\end{equation}
This opens the fascinating possibility that $a(t)$ grows so rapidly that
the integral remains finite as $\Delta t$ goes to infinity. For example,
suppose the scale factor grows like a power of time,
\begin{equation}
a(t) = a_1 \Bigl( \frac{t}{t_1}\Bigr)^b \;\; \Longrightarrow \;\;
H(t) = \frac{b}{t} \;\; \Longrightarrow \;\; q(t) = -1 + \frac1{b} \; .
\label{powerlaw}
\end{equation}
Inflation corresponds to $b > 1$, which is also the case for which the
integral in expression (\ref{masslessET}) remains finite as $\Delta t$
goes to infinity,
\begin{eqnarray}
\lefteqn{a(t) = a_1 \Bigl( \frac{t}{t_1}\Bigr)^b \; , \;\; b > 1}
\nonumber \\
& & \hspace{.5cm} \Longrightarrow \hbar c k \int_{t}^{t + \Delta t} 
\frac{dt'}{a(t')} = \frac{\hbar c k t_1}{a_1 (1 \!-\! b)} \Bigl( 
\frac{t'}{t_1}\Bigr)^{1-b} \Biggl\vert_{t}^{t+\Delta t} 
\longrightarrow -\frac{\hbar c k}{q(t) H(t) a(t)} \; . \qquad
\label{infET}
\end{eqnarray}
Therefore, any massless virtual particle which emerges from empty space 
during inflation with $c k < -q(t) H(t) a(t)$ can persist forever!

\section{Appendix C: How Inflationary Perturbations Fossilize}

Both of gravitons and the simplest kind of massless scalar have the same 
wave function. The wave function $u(t,k)$ for wave number $k$ obeys the 
equations,
\begin{equation}
\ddot{u} + 3 H \dot{u} + \frac{c^2 k^2}{a^2} u = 0 \qquad , \qquad
u \dot{u}^* - \dot{u} u^* = \sqrt{\frac{32 \pi \hbar G}{c}}
\frac{i}{a^3} \; . \label{ueqn}
\end{equation}
Undergraduate physics majors learn to recognize this as the equation of a
harmonic oscillator with a time dependent mass $m(t) = 
\sqrt{\frac{c\hbar}{32\pi G}} \, a^3(t)$ and time dependent characteristic 
frequency $\omega(t,k) = \frac{ck}{a(t)}$. They also learn that the quantum 
mechanical average value of the energy $E(t,k)$ is,
\begin{equation}
E(t,k) = \frac12 m(t) \Bigl[ \vert \dot{u}(t,k) \vert^2 + \omega^2(t,k) 
\vert u(t,k)\vert^2 \Bigr] \; . \label{MMCE}
\end{equation}
Finally, undergraduate physics majors learn how to use expression (\ref{MMCE})
to read off the number $N(t,k)$ of particles which are present at time $t$,
\begin{equation}
E(t,k) = \hbar \omega(t,k) \Bigl[ \frac12 + N(t,k)\Bigr] \; . \label{Ndef}
\end{equation}
The initial factor of $\frac12$ in expression (\ref{Ndef}) is an example
of the irreducible minimum level of excitation implied by the Uncertainty
Principle.

Equation (\ref{ueqn}) cannot be solved for arbitrary deceleration parameter
$q(t)$. However, the data implies $-1 \leq q_i \leq -0.9946$, so it should be
a wonderful approximation to set $q_i = -1$. This means the Hubble parameter
is a constant $H_i$ and the scale factor grows exponentially $a(t) = a_i
{\rm Exp}[H_i (t - t_i)]$. Under these assumptions $u(t,k)$ takes the form,
\begin{equation}
u(t,k) = \sqrt{\frac{\hbar}{2 m(t) \omega(t,k)}} \Bigl[1 + 
\frac{i H_i a(t)}{c k}\Bigr] \, {\rm Exp}\Bigl[-ick \int_{t_i}^t 
\frac{dt'}{a(t')}\Bigr] \; . \label{MMCmode}
\end{equation}
Substituting this solution (\ref{MMCmode}) into the energy (\ref{MMCE}) and
comparing with expression (\ref{Ndef}) is how one derives expression 
(\ref{dSN}) in the main text. If we note that $m(t) \omega(t) = {\rm const} 
\times a^2(t)$ it is easy to show that $u(t,k)$ approaches a constant at 
late times, which is how one derives expression (\ref{fossil}) in the main
text.

Before going any further it is useful to compare the results for $u(t,k)$
with those for conformally coupled particles $c(t,k)$. These wave functions 
can be solved for any scale factor $a(t)$,
\begin{equation}
c(t,k) = \sqrt{\frac{\hbar}{2 m(t) \omega(t,k)}} \,
{\rm Exp}\Bigl[-ick \int_{t_i}^t \frac{dt'}{a(t')}\Bigr] \; . 
\label{MCCmode}
\end{equation}
As it must, this explicit result shows all the features we inferred in
the previous subsection on general grounds. For example, (\ref{infET}) 
resides in the complex exponential, and expression (\ref{conformal}) 
appears in the overall multiplicative factor $1/\sqrt{m(t) \omega(t,k)}
\sim 1/a(t)$. The average energy in this system turns out to be just 
$\frac12 \hbar \omega(t,k)$, so the number of particles created is zero.
That is only the average energy, and scales during primordial inflation
are so enormous that there are still significant quantum gravitational
effects, however, the late time limit of this wave function is zero,
\begin{equation}
\lim_{t \rightarrow \infty} c(t,k) = 0 \; .
\end{equation}
This means that nothing can survive to be observed at later times.

\end{document}